\documentclass[journal,twocolumn]{IEEEoj}

\usepackage[T1]{fontenc}
\usepackage[utf8]{inputenc}
\usepackage{amsmath} 
\usepackage{amsfonts} 
\usepackage{amssymb} 
\usepackage{amsthm}
\usepackage{array}
\usepackage{color}
\usepackage{url}
\usepackage{setspace} 
\usepackage{verbatim}
\usepackage{placeins}
\usepackage[linesnumbered,ruled]{algorithm2e}
\usepackage{multirow}
\usepackage{subcaption}
\usepackage{graphicx}
\usepackage{cite}
\usepackage{stackengine}
\usepackage{booktabs}
\usepackage{multirow}
\usepackage{tablefootnote}
\usepackage{textcomp}

\def\BibTeX{{\rm B\kern-.05em{\sc i\kern-.025em b}\kern-.08em
    T\kern-.1667em\lower.7ex\hbox{E}\kern-.125emX}}
\AtBeginDocument{\definecolor{ojcolor}{cmyk}{0.93,0.59,0.15,0.02}}

\renewcommand{\vec}[1]{{\bf{#1}}} 
 
\newcommand{\tran}{^{\mbox{\scriptsize T}}}
\newcommand{\herm}{^{\mbox{\scriptsize H}}}
\newcommand{\conj}{^{\mbox{\scriptsize *}}}
\newcommand{\fronorm}[3][2]{\Vert #2 - #3\Vert_F^2}
\newcommand{\fro}[1]{\Vert #1\Vert_\textit{F}}
\newcommand{\norm}[1]{\Vert #1\Vert}

\DeclareMathOperator*{\argmin}{arg\,min} 
\newcommand{\delequal}{\mathrel{\ensurestackMath{\stackon[1pt]{=}{\scriptstyle\Delta}}}}

\newcommand{\trace}[1]{\mathrm{tr}\left(#1\right)}

\newcommand{\phib}{\boldsymbol{\phi}}

\newcommand{\psib}{\boldsymbol{\psi}}

\newcommand{\overbar}[1]{\mkern 1.5mu\overline{\mkern-1.5mu#1\mkern-1.5mu}\mkern 1.5mu}

\newcommand{\mybibliography}{\bibliography{jour_short,conf_short,References.bib}}

\title{Channel Charting Aided Pilot Reuse for Massive MIMO Systems with Spatially Correlated Channels}

\author{

LUCAS RIBEIRO\authorrefmark{1}, MARKUS LEINONEN\authorrefmark{1},
MEMBER, IEEE, HANAN AL-TOUS\authorrefmark{2}, OLAV TIRKKONEN\authorrefmark{2}, SENIOR MEMBER, IEEE, MARKKU JUNTTI\authorrefmark{1},
FELLOW, IEEE}

\affil{Centre for Wireless Communications, FI-90014, University of Oulu, Finland}
\affil{Department of Communications and Networking, Aalto University, Finland}
\corresp{CORRESPONDING AUTHOR: Lucas Ribeiro (e-mail: lucas.ribeiro@oulu.fi).}
\authornote{This research has been financially supported by the Academy of Finland under the projects ROHM and 6G Flagship.
The work of M. Leinonen has also been financially supported in part by Infotech Oulu and the Academy of Finland (grant 340171 and 323698).
Preliminary results of this paper were presented in 2020 IEEE Globecom Workshops and 2021 IEEE 32nd Annual International Symposium on Personal, Indoor and Mobile Radio Communications (PIMRC).
}

\begin{document}
\receiveddate{18 October, 2022}
\accepteddate{19 November, 2022}
\doiinfo{OJCOMS.2022.3225054}
\sloppy

\begin{abstract}
Massive multiple-input multiple-output (mMIMO) technology is a way to increase spectral efficiency and provide access to the Internet of things (IoT) and machine-type communication (MTC) devices. To exploit the benefits of large antenna arrays, accurate channel estimation through pilot signals is needed. Massive IoT and MTC systems cannot avoid pilot reuse because of the enormous numbers of connected devices. We propose a pilot reuse algorithm based on channel charting (CC) to mitigate pilot contamination in a multi-sector single-cell mMIMO system having spatially correlated channels. We show that after creating an interference map via CC, a simple strategy to allocate the pilot sequences can be implemented. The simulation results show that the CC-based pilot reuse strategy improves channel estimation accuracy, which subsequently improves the symbol detection performance and increases the spectral efficiency compared to other existing schemes. Moreover, the performance of the CC pilot assignment method approaches that of exhaustive search pilot assignment for small network setups.
\end{abstract}


\maketitle

\section{INTRODUCTION}
\IEEEPARstart{M}{assive} multiple-input multiple-output (mMIMO) technology is a key enabler for the internet of things (IoT) and massive machine-type communication (mMTC) systems~\cite{Chen2021,Callebaut2021,Yan2020,Lee2020,Senel2018}. It supports the access of huge numbers of connected devices through the spatial multiplexing of the user equipments (UEs). Furthermore, the effects of small-scale fading and uncorrelated noise have been shown to asymptotically vanish in mMIMO, significantly improving spectral and energy efficiencies for such systems~\cite{Marzetta2010,Hoydis2013,Ngo2013}. However, to exploit spatial multiplexing and the benefits of channel hardening, accurate channel estimation through pilot signals is needed.

IoT and mMTC systems are characterized by sporadic UE activity, uplink-dominated transmissions, and small packets~\cite{Yan2020,Bockelmann2016,Senel2018}. The design of access techniques to support the huge numbers of such devices is still an open challenge~\cite{Cheng2021}. As opposed to conventional networks connecting mostly human users, the large number of UEs in mMIMO systems precludes the allocation of orthogonal pilot sequences to all of them~\cite{Senel2018,Liu2018}.
Different approaches such as the intelligent reuse of orthogonal pilot sequences and the use of non-orthogonal pilot sequences have been proposed to cope with the shortcoming of orthogonal resources. The first approach is the subject of this work.
A related pilot sequence allocation problem appears also in non-orthogonal multiple-access (NOMA) systems~\cite{Le2021} for which our approach can bring new insights as well.

\textit{Pilot reuse} can be employed within the same cell to acquire the required channel state information (CSI) and avoid excessive signaling, overcoming the lack of resources in mMIMO systems at the cost of \emph{pilot contamination} known also as \emph{pilot collision}.
The contamination or collision of pilot sequences is the interference caused by reusing the same pilot sequence across different UEs to estimate the uplink channels, and it typically affects neighboring cells of a cellular network~\cite{Larsson2014}.
Particularly, when the number of antennas tends towards infinity, the ultimate limiting phenomenon affecting the performance of mMIMO systems is the pilot contamination~\cite{Marzetta2010}.
To tackle the pilot contamination and improve both spectral and energy efficiency, several approaches exploiting the  propagation environment through spatial multiplexing have been proposed~\cite{Bjornson2016,Carvalho2017}.

\subsection{RELATED WORK}
To address the shortage of resources and mitigate the pilot contamination problem in multi-user mMIMO systems, superimposed pilots have been studied in~\cite{Jing2018,Lago2020}. Instead of time-division multiplexing of pilot symbols to estimate the channels and data, pilots are sent to the base station (BS) along with the data, not in dedicated time slots. In~\cite{Lago2020}, Lago \textit{et al.} proposed a mixed approach where first a set of orthogonal pilot sequences is sent, to latter improve the accuracy of the channel estimation, 
and then superimposed pilot symbols are sent to estimate the uplink channels. The results showed that this approach can improve spectral efficiency as compared to purely superimposed or orthogonal pilot schemes.

The design of pilot sequences is considered in~\cite{Van_Chien2018} to mitigate pilot contamination. Van~Chien~\textit{et al.} proposed a framework where the pilots are linear combinations of orthogonal pilot sequences acting as basis vectors, and the resulting non-combinatorial optimization problem aims to find suitable weight coefficients for the basis vectors. 
Xu~\textit{et al.} \cite{Xu2019} utilized deep learning to find such weight coefficients. They formulated a pilot power loading problem to minimize the sum mean square error (MSE) of channel estimation. They have shown that the proposed method improves the sum MSE as compared to other existing methods by employing a deep neural network to learn the MSE-minimizing optimal power allocation to each pilot symbol.

In addition to the aforementioned approaches, the pilot contamination can be alleviated by exploiting the spatial correlation of channels, as presented in~\cite{Sanguinetti2020}. Muppirisetty \textit{et al.} \cite{Muppirisetty2015} proposed a location-aided pilot allocation algorithm, which assumes the knowledge of UEs' physical locations and distribution of the scatterers to estimate the covariance matrices and, subsequently, to suppress pilot contamination. They consider an integer linear problem to assign the pilots. Their results show that the normalized channel estimation error approaches the interference-free scenario. A similar approach, yet using angle-of-arrival (AoA) information for allocating the pilot sequences was proposed in~\cite{Li2018}. Therein, first the UEs are grouped within the cells to alleviate intra-cell pilot contamination. UEs in the same group cannot use the same pilot sequence. 
Next, an algorithm matches the groups with strong potential interference in the neighboring cells into a group collection. Then, a pilot allocation algorithm assigns orthogonal pilot sequences to UEs with strong inter-cell interference. The potential mutual interference is measured on an undirected weighted graph, with vertices representing the UEs in each collection and the edges representing the interference.
The main practical limitation of such location-aware pilot assignment methods is the requirement of knowledge about UEs' physical locations.

Another class of solutions, which do not require the explicit knowledge of UEs' positions, utilizing the spatial correlation to avoid pilot contamination includes \cite{Zhu2015,Dao2018,Ahmed2019,You2015}. In~\cite{You2015}, a pilot reuse strategy called statistic greedy pilot scheduling (SGPS), based on the second-order statistics of the channels, was proposed aiming at minimizing the MSE for both channel estimation and signal detection. The SGPS algorithm seeks to assign orthogonal pilot sequences to UEs with similar channel covariance matrices, according to a metric based on the covariance matrix distance (CMD)~\cite{Herdin2005}, resulting in significant improvement in terms of spectral efficiency as compared to the orthogonal pilot allocation without reuse. 

Nguyen \textit{et al.}~\cite{Nguyen2021} proposed to maximize the minimum weighted sum spectral efficiency by applying a pilot allocation algorithm and a transmission power control policy, aiming to diminish pilot contamination and to mitigate interference for the data transmission, respectively. In~\cite{Ma2019}, Ma~\textit{et al.} utilize sparse Bayesian learning to estimate the channels' spatial signatures. Then, in the channel training phase, UEs with non-overlapping spatial signatures are grouped and assigned with the same pilot sequence, mitigating the overall pilot contamination.

Recently, we proposed a pilot reuse scheme in~\cite{Ribeiro2020,Ribeiro2021} that utilizes \textit{channel charting} (CC\footnote{CC is interchangeably used for ``channel charting'' and ``channel chart'' throughout the paper.}) to exploit the spatial information existing in the CSI to allocate and reuse orthogonal pilot sequences so as to alleviate pilot contamination. CC is a framework proposed in~\cite{Studer2018} to estimate the relative positions of devices in an unsupervised manner. The CC maps the information obtained from the measured long-term CSI at the BS into a low-dimensional chart, in which the relative positions of UEs are preserved. The multi-cell extension of~\cite{Ribeiro2020,Ribeiro2021} is considered in~\cite{Ribeiro2022}. Our results~\cite{Ribeiro2020, Ribeiro2021,Ribeiro2022} show that CC can be utilized to mitigate pilot contamination and improve the channel estimation accuracy.  

\subsection{CONTRIBUTIONS}
In this paper, we address the problem of pilot assignment in a single-cell multi-sector mMIMO system with spatially correlated channels
by employing pilot reuse of a set of orthogonal sequences.
The interference in mMIMO systems is often related to the spatial configuration of UEs, especially to their angular position in relation to the BS. Therefore, a description of the whole azimuth plane is very important to characterize the interference among UEs. 
As directional antennas are usually employed in cellular systems, a multi-sector deployment is needed to provide coverage to the whole angular domain. Furthermore, the multi-sector arrangement allows us to get rid of the mirror angle phenomenon presented in uniform linear arrays (ULAs)~\cite[p. 240]{Bjornson2017} or uniform planar arrays (UPAs).

We formulate a pilot allocation optimization problem that aims at minimizing pilot contamination by exploiting the spatial correlation of the UEs' channels through their second-order statistics. The formulated problem is, in general, hard to solve due to its combinatorial nature, which makes the complexity grow exponentially with the number of UEs and linearly with the number of available pilot sequences. To this end, we propose a practical solution based on CC to create a map that reveals the amount of interference between the UEs by mapping those with strong channel correlation closer than those with weak one. We show through simulation experiments that a simple pilot allocation strategy can be employed if an appropriate interference map is designed. 
Numerical results show that the intelligent CC-based pilot reuse method mitigates pilot contamination and greatly increases spectral efficiency. In a toy example, for a small network setup with few UEs deployed, the proposed method approaches the performance of exhaustive search.

The main \textbf{contributions} of this paper are summarized as follows:
\begin{itemize}
    \item We formulate a pilot allocation problem to optimally reuse a pool of orthogonal pilots in single-cell multi-sector mMIMO systems with spatially correlated channels.
    \item We propose a CC method that translates the amount of interference between the UEs into a low-dimensional chart, creating an interference mapping. The higher the potential interference between two UEs, the closer they reside on the map.
    \item We propose to improve the interference map generated by CC by allowing a variable number of dimensions for CC. The extra degrees of freedom in CC allows us to construct more reliable interference mappings.
    \item We develop a CC-based greedy pilot allocation algorithm that exploits the spatial correlation of the UEs, embedded in the interference mapping. The pilot assignment algorithm aims at maximizing the distances between the UEs on the interference map that use the same pilot sequence to mitigate pilot contamination.
    \item Simulation results show that the proposed pilot allocation can cope with pilot contamination, i.e., it increases channel estimation accuracy and uplink achievable rate, or equivalently, reduces the symbol error rate.
\end{itemize}

The use of second-order statistical information to allocate the pilot sequences or to retrieve UEs' spatial information is not a completely new idea. In~\cite{You2015}, the covariance matrices were fed to a pilot allocation algorithm that computes a CMD-like feature and greedily assigns orthogonal pilot sequences, trying to minimize the interference with the set of UEs previously addressed. 
However, the novelty in our work lies in the design of an UE-interference map of the radio environment through CC. The UE-interference map reveals the big picture of the interference in the cell, which allows employing a low-complexity pilot allocation strategy to mitigate pilot contamination.
 
The present paper builds on our initial works~\cite{Ribeiro2020, Ribeiro2021}. 
Herein, we take a more elaborate approach to the studied pilot contamination problem. We start by deriving
an objective function to assess the overall amount of pilot contamination in the network, which works as a proxy function for the end performance metrics. Using this objective function, we approach the pilot contamination problem and the associated pilot reuse strategy by formulating a rigorous optimization problem that aims to minimize the amount of pilot contamination. This optimization problem is subsequently used as a basis to utilize meta-heuristics and propose the CC-based pilot reuse algorithm. Furthermore, to increase robustness of the proposed method to the changes in the UEs' density, here we propose the use of an adaptive number of dimensions for CC.

\subsection{PAPER OUTLINE}
The rest of the paper is organized as follows. Section~\ref{sec:SystemModel} presents the system model for the multi-sector single-cell 
communication system. The channel estimation and the problem formulation are presented in Section~\ref{sec:CEandPF}. The CC framework to retrieve a radio interference map is proposed in Section~\ref{sec:CC}. In Section~\ref{sec:pilotStrategy}, the proposed pilot allocation strategy utilizing the CC-based interference mapping is devised. The simulation results are presented and discussed in Section~\ref{sec:results}, and Section~\ref{sec:conclusion} concludes the paper.

\textit{Notation:} Boldface lowercase letters, $\vec{x}$, denote column vectors and boldface uppercase letters, $\vec{X}$, denote matrices. The superscripts $(\cdot)\tran$, $(\cdot)\conj$, and $(\cdot)\herm$ denote transpose, conjugate, and conjugate transpose, respectively. The average value of $\vec{x}$ is $\Bar{\vec{x}}$. The multivariate circularly symmetric complex Gaussian distribution with covariance matrix $\vec{R}$ and mean $\vec{x}$ is denoted as $\mathcal{CN}(\vec{x},\vec{R})$. The $n\times n$ identity matrix is presented as $\vec{I}_n$. The vector of all entries $1$ is denoted as $\vec{1}$. The element-wise product between ${\vec{A}}$ and ${\vec{B}}$ is expressed as ${\vec{A}\odot\vec{B}}$. The expected value of $\vec{x}$ is denoted as $\mathbb{E}[\vec{x}]$. The ${\ell_2\text{-norm}}$ and the Frobenius norm are denoted as $\Vert\cdot\Vert_2$ and $\fro{\cdot}$, respectively. The trace of ${\vec{X}}$ is denoted as ${\trace{\vec{X}}}$.

\section{SYSTEM MODEL}
\label{sec:SystemModel}

\begin{figure}[t]
    \centering
    \includegraphics[scale=0.3]{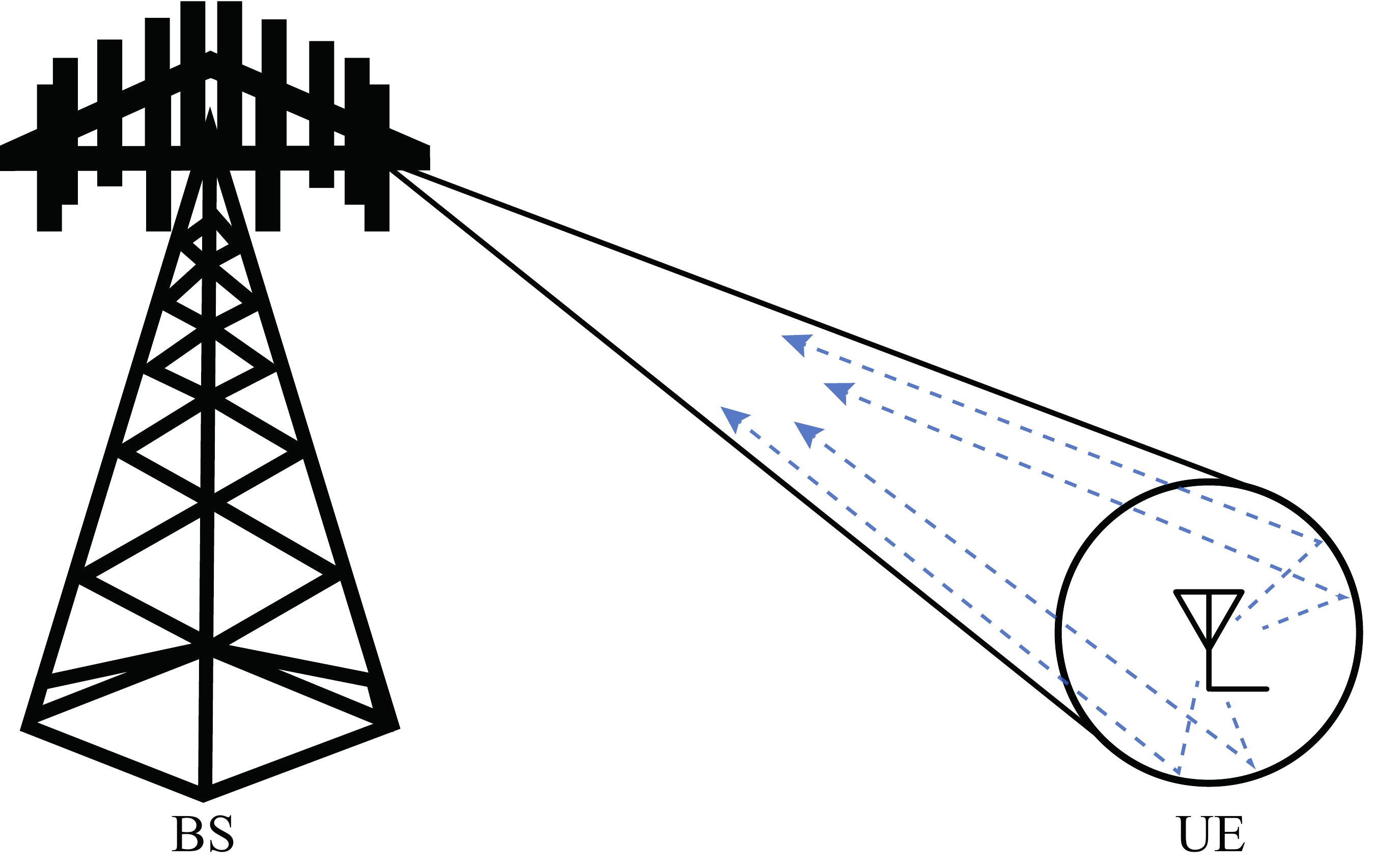}
    \caption{The uplink channel between a single-antenna UE and a BS equipped with ${S=3}$ ${M\text{-element}}$ ULAs. The multi-path components are concentrated in a ring around the UEs.}
    \label{fig:channel}\vspace{-12pt}
\end{figure}

We consider a mMIMO uplink communication scheme with a set ${\mathcal{N}=\{1,\ldots,N\}}$ of $N$ single-antenna UEs uniformly distributed within a cell, from which only ${K<N}$ are randomly active at any given time. In order to focus on the development of pilot assignment strategies to improve the channel estimation accuracy, and, subsequently, enhance the system performance, we assume that the BS knows the set of active UEs at each transmission instant. This assumption can be invoked by the fact that there exists a multitude of compressed sensing based approaches, e.g., \cite{chen2018sparse,Senel2018,Djelouat2020Joint}, to detect the set of active UEs, and thus, we leave it outside of the scope of this paper.

The cell is divided into $S$ sectors\footnote{In practical systems, the outdoors coverage area is usually divided into three or six sectors of ${120^\circ}$ or ${60^\circ}$~\cite{Morales2019}. We consider $S=3$ in the examples.}. The BS receives the UEs' transmitted signals 
through $S$ ULAs\footnote{The extension to UPAs is straightforward~\cite{Yu2021}.} each having $M$ antenna elements. The array response vector for a ULA is given by
\begin{equation} 
\vec{a}_{\mathrm{r}}(\theta) = \left[{1, e^{-j2\pi \Delta_{\mathrm{r}}\cos(\theta)}, \ldots, e^{-j2\pi (M-1)\Delta_{\mathrm{r}} \cos(\theta)}}\right]\tran, 
\end{equation}
where ${\Delta_{\mathrm{r}}}$ is the normalized spacing between the antenna elements in units of wavelengths and $\theta$ is the AoA, i.e., the incident angle of the received signal on the antenna array\cite[Sec. 7.2.1]{TseBook}. The ULA geometry causes resonance at certain angle pairs, a phenomenon known as mirror angles~\cite[p. 258]{Bjornson2017}.
Multi-sector processing allows the proposed CC-based method to handle the ULAs' mirror angles effect by combining the signal from different arrays, thus, granting coverage to UEs at the whole extent of the angular domain, ${[0,2\pi]}$.

We adopt the one-ring channel model that assumes the multi-path components are concentrated around the UEs while the BS is in an elevated position, thus lacking scatterers in its proximity. Each multipath reaches the antenna array from a particular angle within a ring close to the UE, which makes the channels spatially correlated with independent gains and phase rotations~\cite[p. 236]{Bjornson2017}.
Fig.~\ref{fig:channel} depicts the channel model between a UE and the BS. Accordingly, the uplink channel vector for user ${n\in\mathcal{N}}$ at the ULA of sector ${s\in \mathcal{S}=\{1,\ldots,S\}}$ is modelled as a superposition of $L$ propagation paths as
\begin{equation} 
\label{eq:hk-sector}
\vec{h}_{n,s} = \frac{1}{\sqrt{L}}\sum_{l=1}^{L}\sqrt{\beta_{n,s,l}}\alpha_{n,s,l}\vec{a}_{\mathrm{r}}(\theta_{n,s,l}),
\end{equation}
where ${\alpha_{n,s,l}}$ is the complex gain of the $l$th path, which is assumed to be an independent and identically distributed (i.i.d.) complex Gaussian random variable with zero mean and ${\mathbb{E}[|\alpha_{n,s,l}|^2]=1}$. 
The large-scale propagation effects and the BS antenna gain in the channel are captured in ${\beta_{n,s,l} \in \mathbb{R}}$. Any path loss model can be used to model ${\beta_{n,s,l}}$, here, we consider the free-space path loss model as described in~\cite[Eq. (2.7)]{Goldsmith2005}, which is defined as
 \begin{equation}
 \beta_{n,s,l} \delequal 10^{\frac{G_{\mathrm{A}}(\theta_{n,s,l})}{10}}\left(\frac{\lambda}{4\pi d_{n}}\right)^2, 
 \label{eq:beta}
 \end{equation}
where $\lambda$ is the wavelength and ${d_{n}}$ is the distance between UE $n$ and the BS. In~\eqref{eq:beta}, ${G_{\mathrm{A}}(\theta_{n,s,l})\in \mathbb{R}}$ represents the antenna gain through the $l$th path for user $n$ at the ULA of sector $s$, and is given in dB by~\cite[Table 7.1-1]{TR36873,Morales2019} 
\begin{equation}
    G_{\mathrm{A}}(\theta_{n,s,l}) = G_{\mathrm{A_{max}}} {-\textrm{min}\left[12\left(\frac{\theta_{n,s,l}}{\theta_{3\textrm{dB}}}\right)^2,\;A_\textrm{max}\right]},
    \label{eq:AntennaGain}
\end{equation} 
where ${G_{\mathrm{A_{max}}}}$ is the maximum antenna gain, ${\theta_{3\textrm{dB}}}$ is the half power beamwidth, $A_\textrm{max}$ is the maximum attenuation in dB at the ULAs. Above, ${\theta_{n,s,l}}$ is the AoA of the $l$th path, ${l=1,\ldots,L}$. The AoA for the $L$ paths between UE $n$ and ULA $s$ can be modelled as i.i.d.\ random variables with uniform distribution ${U(\theta_{n,s}^{\mathrm{min}},\theta_{n,s}^{\mathrm{max}})}$, with ${\theta_{n,s}^{\mathrm{min}}=\Bar{\theta}_{n,s}-\sqrt{3}\sigma_{\theta}}$ and ${\theta_{n,s}^{\mathrm{max}}=\Bar{\theta}_{n,s}+\sqrt{3}\sigma_{\theta}}$. Here, ${\Bar{\theta}_{n,s}\in [0,2\pi]}$ is the incident angle between UE $n$ and the ULA of sector $s$, ${\sigma_{\theta}}$ is the angular standard deviation, which specifies the AoA interval, ${\mathcal{A}_{n,s}=[\theta_{n,s}^{\mathrm{min}},\theta_{n,s}^{\mathrm{max}}]}$, for the possible incoming multi-path components arriving from UE $n$ at the ULA of sector $s$.

\begin{figure}[t]
    \centering
    \includegraphics[scale=0.49]{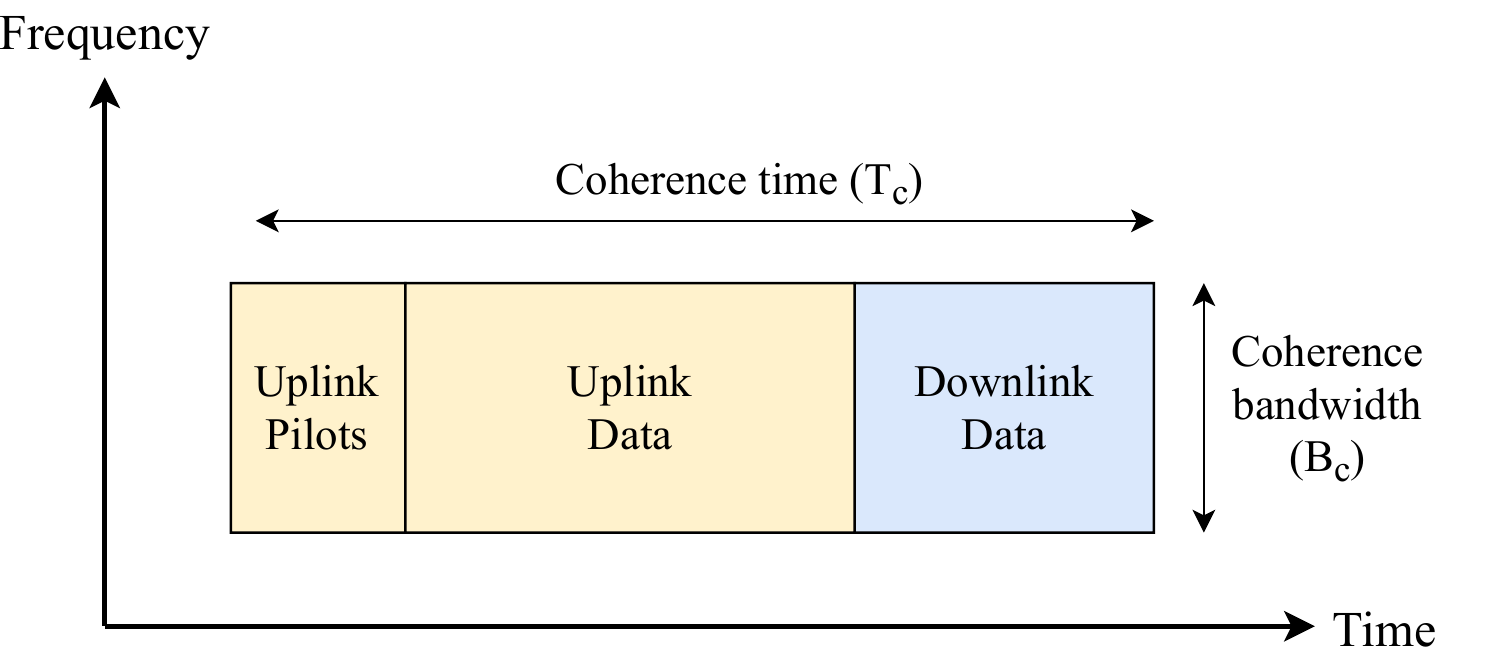}
    \caption{Illustration of transmission phases in the considered TDD system. At the beginning of each coherence block, $\tau$ symbols are sent to estimate the channel.}
    \label{fig:TDD}\vspace{-12pt}
\end{figure}

We consider a time-division duplex (TDD) system, as depicted in Fig.~\ref{fig:TDD}. We assume that during one coherence block, the channels are time-invariant and flat fading. For TDD systems, downlink and uplink channels are symmetric, i.e., the downlink channel can be inferred at the BS from the uplink one~\cite{TseBook}. At each coherence block, the set of active UEs transmit $\tau$ known symbols to the BS for channel estimation. Right after transmitting the pilot symbols, the $K$ active UEs transmit their data to the BS. Then, the remaining time within the coherence block is used for downlink communication.

\section{CHANNEL ESTIMATION AND PROBLEM FORMULATION}
\label{sec:CEandPF}

The UEs are assigned with pilot sequences taken from a pool of $\tau$ \textit{orthogonal} sequences. However, due to a vast number of UEs in mMIMO networks, we assume that ${\tau\ll N}$, so that global \textit{pilot reuse} is employed in the cell area. Consequently, in the channel estimation phase, the same pilot is shared on the average by ${K/\tau}$ UEs. Let ${\mathcal{K}=\{1,\ldots,K\}\subseteq\mathcal{N}}$ represent the set of active UEs at a given transmission interval. Let ${\mathcal{T}=\{1,\ldots,\tau\}}$ be the set of indices of the available pilot sequences. UE ${k\in\mathcal{K}}$ transmits a pilot signal ${{\psib}_k=\sqrt{p_{\mathrm{u}}}\phib_{\pi_k}}$, where ${p_{\mathrm{u}}}$ is the power of each pilot symbol, ${\pi_k\in\mathcal{T}}$ is the index of the pilot sequence assigned to UE $k$, and ${\phib_{\pi_k}\,\in\,\mathbb{C}^{\tau}}$ is the corresponding pilot sequence, with power $\tau$, taken from the orthogonal pilot book ${\vec{\Phi}=[\vec{\phib}_1,\ldots,\vec{\phib}_{\tau}] \in \mathbb{C}^{\tau\times \tau}}$. We define the set of interfering UEs to UE $k$ as ${\mathcal{I}_k}$, i.e., ${\mathcal{I}_k=\{j\mid j\in\mathcal{K}\setminus\{k\}, \, \pi_j=\pi_k\}}$.

\subsection{CHANNEL ESTIMATION}
The compound channel ${\vec{h}_{k}\in\mathbb{C}^{MS}}$
between UE ${k\in\mathcal{K}}$ and the BS is given by
 \begin{equation}
\label{eq:hk}
\vec{h}_{k} =
\begin{bmatrix}
\vec{h}_{k,1} \\
\vdots\\
\vec{h}_{k,S}
\end{bmatrix},
\end{equation}
with covariance matrix ${\vec{R}_{k}=\mathbb{E}[\vec{h}_{k}\vec{h}_{k}\herm]\in\mathbb{C}^{MS\times MS}}$. The linear minimum mean square error (LMMSE) estimator is deployed at the BS to jointly estimate the channel vectors of the active UEs across the $S$ sectors. Here, because the channels are Gaussian distributed, the MMSE estimator reduces to the LMMSE estimator~\cite[Sec. 12.3]{Kay1993}. Importantly, Yin et al.~\cite{Yin2013} showed that the MMSE estimator can recover the channel estimates as in an interference-free scenario given that the covariances' signal subspaces are orthogonal. In our framework, this orthogonality raises from the restricted angles of the multipaths impinging on the receive antenna from each UE according to the one-ring channel model.

We assume that the channel covariance matrices of all UEs (active and inactive ones), i.e., ${\vec{R}_{n}=\mathbb{E}[\vec{h}_{n}\vec{h}_{n}\herm], \,n\in\mathcal{N}}$, are known at the BS. This may be a challenging assumption for medium/high mobility UEs but is still feasible~\cite{Sanguinetti2020}. 
In practice, an initial training phase is required to obtain the first estimates of the covariance matrices. Several techniques to obtain the covariance matrices are explored in~\cite{Sanguinetti2020} and in the references therein. 
For instance, the BS can keep updating each ${\vec{R}_{n}}$ based on the estimated channel, i.e., the BS can approximate the covariance matrices by their sampled versions.
Although a low-mobility scenario is readily applicable for the proposed method (e.g., electricity and water metering, environmental monitoring, building automation, city lightning, etc.), it is not stringently restricted to this scenario. Namely, the only requirement is that accurate channel statistics are available, meaning that for medium/high mobility, the statistics need to be updated more frequently.

The received signal for channel estimation at the BS, ${\vec{Y}\in\mathbb{C}^{MS \times \tau}}$, can be written as 
\begin{equation}
\vec{Y} = \vec{H}\vec{\Check{\Psi}} + \vec{N},
\label{eq:receivedSignal}
\end{equation}
where ${\vec{H}=\left[\vec{h}_{1},\ldots,\vec{h}_{K}\right]\in\mathbb{C}^{MS \times K}}$ is the channel matrix for the active UEs, ${\vec{\Check{\Psi}}=\left[{\psib}_1,\ldots,{\psib}_K\right]\tran\in\mathbb{C}^{K\times \tau}}$ is the pilot signal matrix, and ${\vec{N}=\left[\vec{n}^{1},\ldots,\vec{n}^\tau\right]\in\mathbb{C}^{MS \times \tau}}$ is the noise matrix. We model the noise as an i.i.d. complex Gaussian random variable ${\vec{n} \sim \mathcal{CN}(\vec{0},\sigma_\mathrm{n}^2\vec{I}_{MS})}$, where ${\sigma_\mathrm{n}^2}$ is the noise power at each antenna element.

The LMMSE estimate of the channel between UE $k$ and the BS, ${\vec{h}_{k}}$ in \eqref{eq:hk}, is given as~\cite{You2015}
\begin{equation}
\label{eq:hk_hat}
    \vec{\hat{h}}_k = \vec{R}_{k}\vec{Q}_{k}^{-1}\vec{y}_{k}^{\mathrm{p}},
\end{equation}
where ${\vec{y}_{k}^{\mathrm{p}}\in\mathbb{C}^{MS}}$ represents the processed received signal for UE $k$ after correlating the received signal with the pilot sequence assigned to UE $k$, i.e.,
\begin{equation}
\begin{split}
    \vec{y}_{k}^{\mathrm{p}} &= \frac{1}{p_{\mathrm{u}} \tau}\vec{Y}{\psib}_{k}^{*}, \\
    &=\vec{h}_k+\underbrace{\sum_{j\,\in\, \mathcal{I}_k}{\vec{h}_j}}_\text{Pilot Interference}+\frac{1}{p_{\mathrm{u}}\tau}\vec{N}{\psib}_{k}^{*} ,
\end{split}
\label{eq:LS}
\end{equation}
and ${\vec{Q}_{k}=\mathbb{E}[\vec{y}_{k}^{\mathrm{p}}\left(\vec{y}_{k}^{\mathrm{p}}\right)\herm]\in\mathbb{C}^{MS\times MS}}$ is the covariance matrix of the processed received signal, given as
\begin{equation}
    \vec{Q}_{k}=\vec{R}_k+\sum_{j\in \mathcal{I}_k}\vec{R}_j + \frac{\sigma_\mathrm{n}^2}{p_\mathrm{u}\tau}\vec{I}_{MS}.
    \label{eq:Rerror}
\end{equation}

Due to the orthogonality principle of the MMSE estimator~\cite[Sec. 12.4]{Kay1993}, the channel estimation error is independent of ${\vec{\hat{h}}_{k}}$ and given by ${\vec{\tilde{h}}_{k}=\vec{h}_{k}-\vec{\hat{h}}_{k}\in\mathbb{C}^{MS}}$. Thus, the error covariance matrix for user $k$ is given by~\cite{You2015} 
\begin{equation}
    \vec{R}_{\tilde{\vec{h}}_{k}}=\vec{R}_{k}-\vec{R}_{k}\vec{Q}_{k}^{-1}\vec{R}_{k},
    \label{eq:err-covariance}
\end{equation}
and the corresponding MSE by
\begin{equation}
\begin{split}
\text{MSE}_k &=\trace{\vec{R}_{\vec{\tilde{h}}_k}}\\
&=\trace{\vec{R}_{k}-\vec{R}_{k}\Big({\vec{R}_k+\sum_{j\in \mathcal{I}_k}\vec{R}_j + \frac{\sigma_\mathrm{n}^2}{p_\mathrm{u}\tau}\vec{I}_{MS}}\Big)^{-1}\vec{R}_{k}}.
\end{split}
\label{eq:MSE}
\end{equation}
To obtain an interference-free estimate of ${\vec{h}_k}$, a vector basis for the set of matrices ${\{\vec{R}_j|j\in\mathcal{I}_k\}}$ must be orthonormal to a basis vector for ${\vec{R}_k}$. 
As proven in~\cite{Yin2013}, this condition is fulfilled, for an asymptotic regime (${M\rightarrow\infty}$), if the multipath components of the interfering UEs lay outside the AoA interval for UE $k$, ${\mathcal{A}_{k,s}}$ in all sectors $s$. 
In that case, the interfering signals will fall in the null space of ${\vec{R}_k}$, and the MSE is going to be purely noise-limited. 
Therefore, the AoA intervals -- or equivalently the eigenstructure of the channel covariance matrices -- will determine the impact of the pilot contamination with large-antenna arrays~\cite{Yin2013,You2015}.

\subsection{PROBLEM FORMULATION}
The massive number of devices in mMIMO systems prevents the allocation of unique orthogonal pilot sequences to all UEs, which necessitates pilot reuse. Pilot contamination affects the capability to mitigate interference during the data transmission phase due to the reduced estimation quality and the statistical dependence of the channel estimates~\cite[p. 252]{Bjornson2017}. 
Furthermore, the strongest interference is usually caused by UEs within the same cell~\cite[p. 246]{Bjornson2017}. Therefore, the coordination of pilot assignment becomes essential for intra-cell pilot reuse systems.

The expression in~\eqref{eq:LS} makes the effect of the pilot reuse on the channel estimation accuracy evident. However, we can avoid pilot contamination by assigning orthogonal pilot sequences to the UEs with non-orthogonal channels, i.e., ${\trace{\vec{R}_{n}\herm\vec{R}_{j}}\neq0}$. In practice, it is not likely that the covariance matrices for distinct UEs are fully orthogonal, i.e., that their channels are spatially uncorrelated. Nevertheless, as shown in~\cite{Yin2013}, it is still possible to mitigate pilot contamination between the UEs that have non-overlapping AoA intervals, i.e., ${\mathcal{A}_{k,s}\cap\mathcal{A}_{j,s}=\{\emptyset\}, \forall k\neq j}$, and $\forall s$. 
This behavior is also reported in~\cite[p. 341]{Bjornson2017} as the asymptotic spatial orthogonality between covariance matrices, and it can be written as
\begin{equation}
    \lim_{M \to \infty}{\frac{1}{M}\trace{\vec{R}_{n}\herm\vec{R}_{j}} = 0},\quad\forall\,n\neq j,
   \label{eq:SpatialOthogonality}
\end{equation}
where ${\trace{\vec{R}_{n}\herm\vec{R}_{j}}}$ is the standard inner product on ${\mathbb{C}^{M\times M}}$, measuring the orthogonality between ${\vec{R}_{n}}$ and ${\vec{R}_{j}}$.

Essentially, the pilot allocation algorithm must assign orthogonal pilot sequences to UEs with potentially strong interference. Thus, we can minimize the pilot contamination effect by wisely assigning the pilot sequences such that UEs with overlapping AoA intervals get orthogonal pilot sequences.

As motivated above, the angular separation between two UEs is closely related to their degree of orthogonality. An intuitive measure to assess the UEs' orthogonality level is presented in~\cite{You2015}, based on the inner product of the covariance matrices. Similarly, we define the normalized channel spatial correlation between UEs $n$ and $j$ as
\begin{equation}\label{eq:norm_channel_correlation}
    \delta(\vec{R}_n,\vec{R}_j) = \frac{\trace{\vec{R}_n\herm\vec{R}_j}}{\fro{\vec{R}_n}\fro{\vec{R}_j}}.
\end{equation}
The spatial correlation between two UEs expresses how orthogonal the channels for the pair of UEs are, or how close these UEs are in the angular domain. The higher the correlation between the channel covariance matrices, the closer the UEs are in the angular domain, and the lower is the orthogonality between their channels.

To exploit the spatial characteristics of the channels for large antenna systems presented in~\eqref{eq:SpatialOthogonality}, we formulate the pilot assignment problem to suppress the interference from UEs that share the same second-order statistics. More precisely, when UEs $n$ and $j$ share a pilot, we measure their contribution to the networks' pilot contamination by ${\delta(\vec{R}_n,\vec{R}_j)}$; otherwise, the contribution is zero. Thus, we aim at minimizing the total amount of pilot contamination in the network through the optimization problem
\begin{equation}
\begin{aligned}
\underset{\{\pi_{n}\in\mathcal{T}\}_{n=1}^{N}}{\textrm{minimize}} \quad & \frac{1}{\tau N(N-1)/2}\sum_{n\in\mathcal{N}}\,\sum_{j>n}\delta(\vec{R}_n,\vec{R}_j){\phib_{\pi_n}\tran\phib_{\pi_j}},
\end{aligned}
\label{eq:assignment-problem}
\end{equation}
where the optimization is over the UE-to-pilot index mapping, i.e., over the $N$ integer variables ${\pi_{n}}$, each taking values on $\{1,\ldots,\tau\}$. The pre-summation factor normalizes the effect of changing the number of UEs and the length of the pilot sequences, but do not affect the minimization problem. Note that, due to orthogonality of the pilot codebook, the inner product ${\phib_{\pi_n}\tran\phib_{\pi_j}}$ acts as an indicator function taking the value $\tau$ if ${{\pi_n}={\pi_j}}$ and $0$ otherwise. Thus, the cost per UE pair ${(n,j)}$ is always either ${\delta(\vec{R}_n,\vec{R}_j)}$ or $0$. With this logic, to minimize~\eqref{eq:assignment-problem}, one should assign orthogonal pilots, distinct indices, to UEs with strong spatial correlation. It is also worth noting that the normalization of the inner product in~\eqref{eq:norm_channel_correlation} ensures fair pilot assignment for UEs with poor channels, as the interference level is normalized for each pair of UEs.

Problem \eqref{eq:assignment-problem} can be understood as a vertex coloring problem on an edge-weighted graph, and as such it is NP-hard~\cite{Kolen1996}. For small-to-moderate-sized networks, the optimal pilot assignment can be found via exhaustive search, but for larger networks, sub-optimal algorithms have to be used. From~\cite{Vredeveld2003}, it is known that solving weighted graph problems with simplified local search algorithms can lead to far-from-optimal solutions. Motivated by this, we search for practical, yet efficient metaheuristics to allocate pilot sequences. We devise a two-stage method to find a practical solution to \eqref{eq:assignment-problem} by the following procedure. The first stage utilizes CC to retrieve an interference map of the UEs, determined by the coefficients ${\delta(\vec{R}_n,\vec{R}_j)}$ in \eqref{eq:norm_channel_correlation}. In particular, we tailor the underlying interference pattern into a lower dimensional map so that this map subsequently allows performing a simple greedy pilot assignment strategy conforming to the structure of the optimal solution of \eqref{eq:assignment-problem}: assign orthogonal pilot sequences to UEs with strong pair-wise interference. In what follows, the proposed interference mapping is presented in Section \ref{sec:CC}, and the proposed pilot assignment method is presented in Section \ref{sec:pilotStrategy}.

\section{CHANNEL CHARTING TO RECOVER AN INTERFERENCE MAP}
\label{sec:CC}

\begin{figure*}[t]
    \centering
    \begin{subfigure}{0.45\textwidth}
    \includegraphics[width=\textwidth]{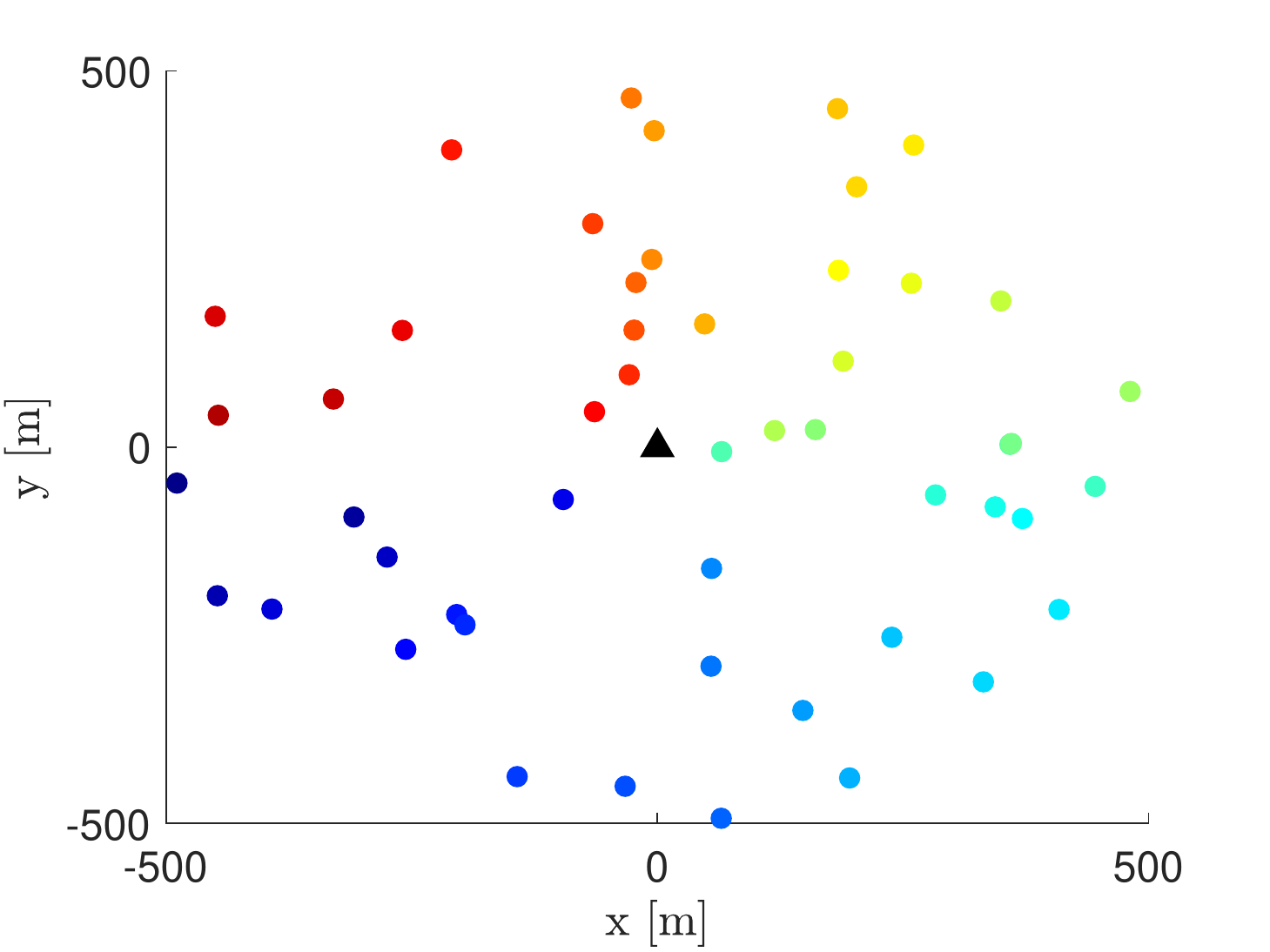}
    \caption{User deployment.}
    \label{subfig:users}
    \end{subfigure}
    \begin{subfigure}{0.45\textwidth}
    \includegraphics[width=\textwidth]{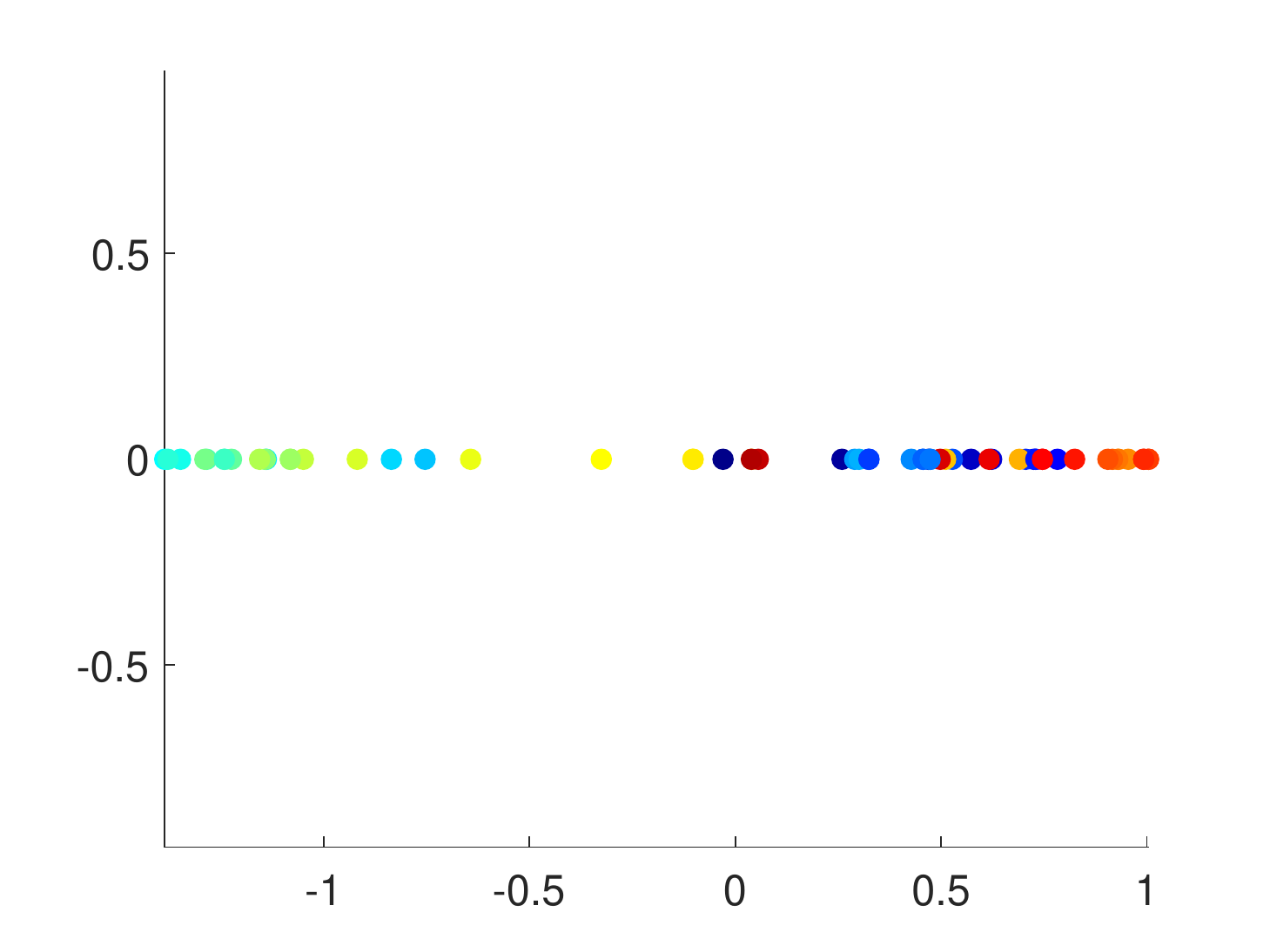}
    \caption{1D CC.}
    \label{subfig:CC40UEs1D}
    \end{subfigure}\\
    \begin{subfigure}{0.45\textwidth}
    \includegraphics[width=\textwidth]{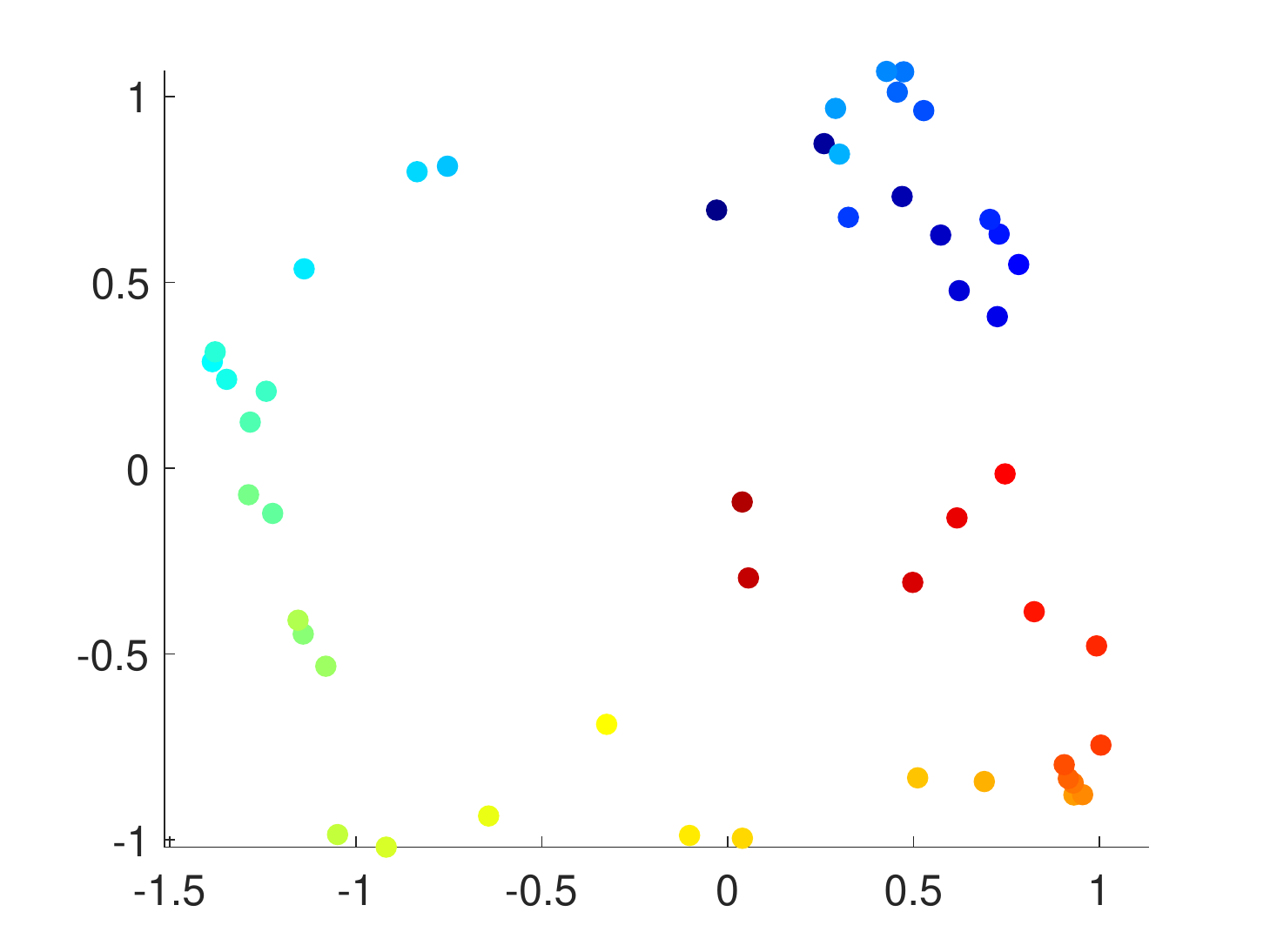}
    \caption{2D CC.}
    \label{subfig:CC40UEs2D}
    \end{subfigure}
    \begin{subfigure}{0.45\textwidth}
    \includegraphics[width=\textwidth]{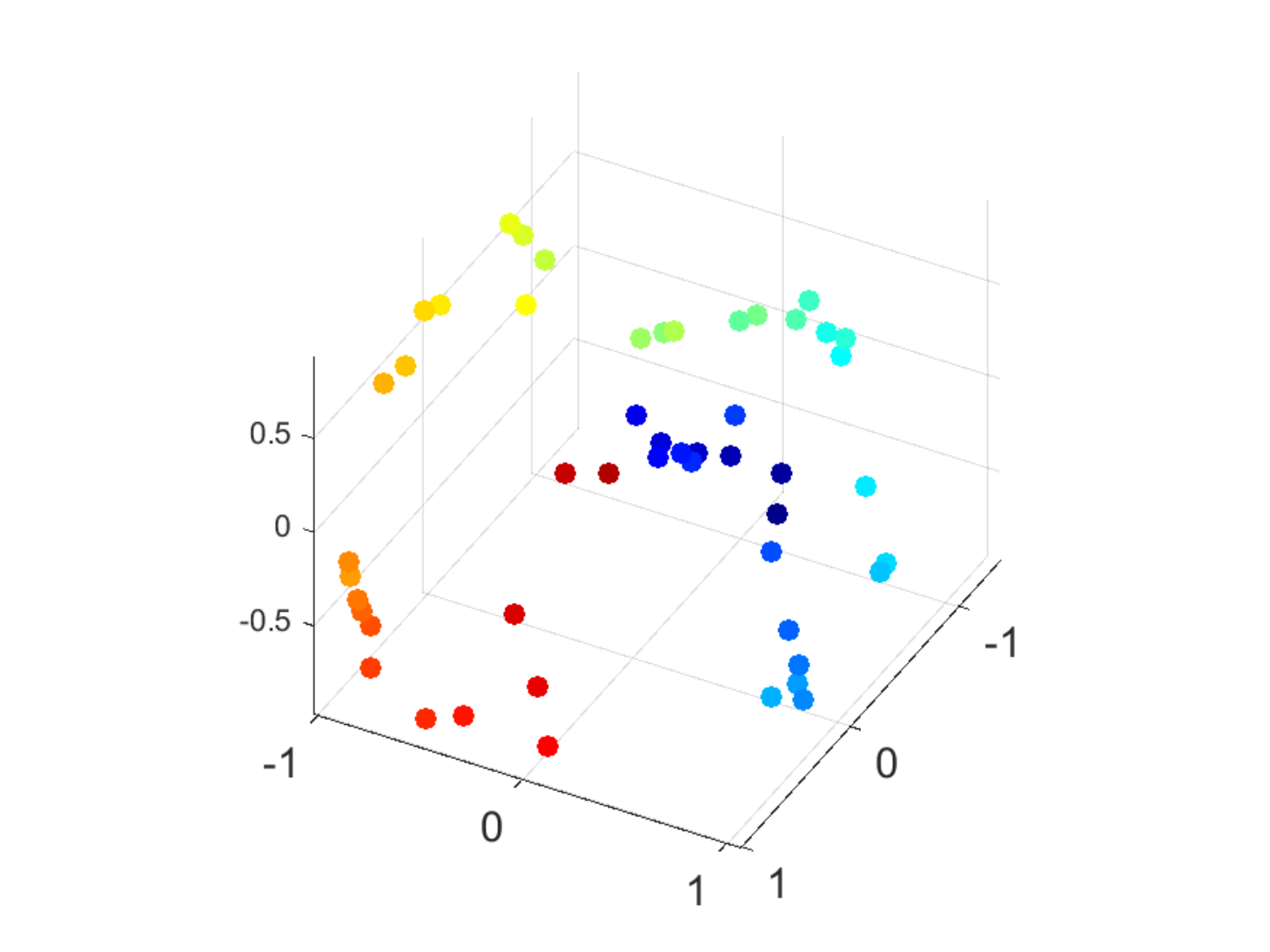}
    \caption{3D CC.}
    \label{subfig:CC40UEs3D}
    \end{subfigure}
    \caption{Channel charting mapping using Isomap. (a) shows the physical location of 40 UEs colored based on their AoA to the BS. The BS, represented by the black triangle, is equipped with ${S=3}$ ULAs, each having ${M=64}$ antenna elements. (b)--(d) represent the CC mapping of (a), for ${C=1}$, ${C=2}$ and ${C=3}$ dimensions, respectively.}
    \label{fig:CC}
\end{figure*}

In this section, we introduce a CC method to retrieve a UE-interference map from the information embedded in the coefficients ${\delta(\vec{R}_n,\vec{R}_j)}$ in \eqref{eq:norm_channel_correlation}. Recall from above that this represents the first (and crucial) stage of our proposed solution to problem \eqref{eq:assignment-problem}: the generated CC-based interference map will be the main input to the second stage, i.e., for the actual pilot assignment algorithm presented in Section~\ref{sec:pilotStrategy}.

The standard CC, as presented in~\cite{Studer2018}, aims at reconstructing a low-dimensional embedding from the CSI features so as to be a good representation of the spatial geometry of UEs. Here, we want to create a map that preserves the angular neighborhood in a low-dimensional chart, building on the fact that the angular separation between UEs represents their inter-UE interference, which is measured by~\eqref{eq:norm_channel_correlation}. The target is to map spatially correlated UEs, with potentially strong interference, close in the interference map.

Clearly, the features to generate the interference mapping must contain spatial information about the UEs, particularly, their angular positions.
To this end, the second-order statistics of CSI, i.e., the covariance matrices ${\vec{R}_n,\,n\in\mathcal{N}}$, available at the BS are good candidates to create CC map. Beyond carrying the required information, they possess desirable characteristics such as varying over larger time scales as compared to the instantaneous CSI. However, if one cannot properly estimate the channel covariance matrices, the UE-interference map may not reflect the true potential interference between UEs, which degrades the performance of the proposed method. Rapid and abrupt changes in the channel geometry may cause significant variations in the covariance matrices. Such cases are not in the scope of the current paper, but form an important item for further study.

The CMD metric~\cite{Herdin2005} measures the normalized orthogonality between two covariance matrices, which makes it a good candidate as a feature for the optimization problem in~\eqref{eq:assignment-problem}. Accordingly, the ${(n,j)\text{-th}}$ element of the feature matrix ${\vec{F}\in\mathbb{R}^{N\times N}}$ generated from the covariance matrices of UEs $n$ and $j$ is
\begin{equation}\label{eq:CMD_centralized}
    [\vec{F}]_{n,j}=1-\delta(\vec{R}_n,\vec{R}_j),
\end{equation}
where ${\delta(\cdot,\cdot)}$ is the normalized channel spatial correlation of~\eqref{eq:norm_channel_correlation}. The more the signal spaces of $\vec{R}_n$ and $\vec{R}_j$ overlap, the larger the interference between UE $n$ and $j$, and the smaller is the distance measured by~\eqref{eq:CMD_centralized}~\cite{Herdin2005}. Because of its normalization factor, it can also be interpreted as an angular distance between the UEs. The $n$-th column of ${\vec{F}=[\vec{f}_1,\ldots,\vec{f}_N]\in\mathbb{R}^{N\times N}}$, ${\vec{f}_n\in\mathbb{R}^N}$, represents the feature or dissimilarity vector that contains the distance of UE $n$ to all $N$ UEs, including itself.

After acquiring the feature matrix $\vec{F}$, the last step to get CC is to apply a function $\mathcal{C}$ on ${\vec{f}_n,\; \,n\in\mathcal{N}}$, to map the set of
UE features to a lower dimensional chart,
\begin{equation}
  \mathcal{C}:\vec{f}_n\mapsto\vec{z}_n,  
  \label{eq:cc-mapping}
\end{equation}
where ${\vec{z}_n\in\mathbb{R}^{C}}$ is the point in the $C$-dimensional CC corresponding to feature ${\vec{f}_n}$, where typically ${C\ll{N}}$.
Apart from providing a lower dimensional representation of the mutual interference between UEs, which may help to visualize an interference pattern, the dimensionality reduction (DR) technique maps the features to the dimensions that have the most relevant attributes to characterize the interference. The DR technique is essential to alleviate the \emph{distance concentration} effect. This phenomenon reduces the difference in distance between near and far neighbors as the features' dimensionality increases, causing problems for algorithms that rely on nearest neighbor search -- the category in which our pilot allocation algorithm falls into~\cite{Beyer1999,Zimek2012}. Therefore, DR techniques will be applied, as described next.

\subsection{DIMENSIONALITY REDUCTION TECHNIQUES}
\label{subsec:DR}
Ultimately, we wish to design a function $\mathcal{C}$ in \eqref{eq:cc-mapping} that maps the feature vectors ${\vec{f}_n,\, n\in\mathcal{N}}$, into a lower dimensional chart such that similar features are mapped close while highly dissimilar features are mapped far away in the chart. 
Several DR techniques have been used to generate CCs~\cite{Studer2018,Lei2019,Huang2019,Agostini2020,Ribeiro2020,Ribeiro2021}. Here, two DR techniques are briefly presented: 1) principal component analysis (PCA) and 2) Isomap, which will be the DR technique used in our numerical experiments. PCA is a very robust and easy-to-implement algorithm, as it performs a linear mapping from the feature domain to CC, which, in turn, may be a limiting factor for non-linear embeddings. On the other hand, Isomap can handle more complex scenarios because it is not restricted to linear mappings.\vspace{-4mm}

\subsubsection{Principal Component Analysis} PCA is a widely used algorithm, which performs a linear projection of high-dimensional data onto a subspace of lower dimension~\cite{Bishop2006}. The main idea behind PCA is to find the components that maximize the variance of the projected data. Let ${\vec{C}=\vec{I}_{N}-\frac{1}{N}\vec{1}\vec{1}\tran\in\mathbb{R}^{N\times N}}$ be the centering matrix. An interference map is then formed by applying singular value decomposition on the covariance of the centered feature matrix, ${\vec{K}=\vec{F}\tran\vec{C}\vec{F}\in\mathbb{R}^{N\times N}}$, and retrieving the coordinates in the ${C\text{-dimensional}}$ space with the largest variance values implying the maximum  interference~\cite{Williams2002}. Thereby, we obtain the CC points, ${\vec{Z}_{C}=\left[\vec{z}_1,\ldots,\vec{z}_N\right]\in\mathbb{R}^{C\times N}}$, via PCA as
\begin{equation}
    \vec{Z}^{\mathrm{PCA}}_{C}=\big[\sqrt{\lambda_1}\vec{u}_1,\ldots,\sqrt{\lambda_{C}}\vec{u}_{C}\big]\tran,
    \label{eq:CC-PCA}
\end{equation}
where ${\lambda_1,\lambda_2,\ldots,\lambda_C}$ are the $C$ largest eigenvalues of ${\vec{K}}$ and ${\vec{u}_1,\vec{u}_2,\ldots,\vec{u}_C}$ are the associated eigenvectors.

Although PCA has shown good performance for single-sector scenarios~\cite{Ribeiro2020}, its linearity reduces its capability to achieve good mapping in more challenging scenarios, like the 3-sector deployment considered herein. Thereby, more sophisticated DR techniques are required.

\begin{figure*}[t]
    \centering
    \begin{subfigure}{0.45\textwidth}
    \includegraphics[width=\textwidth]{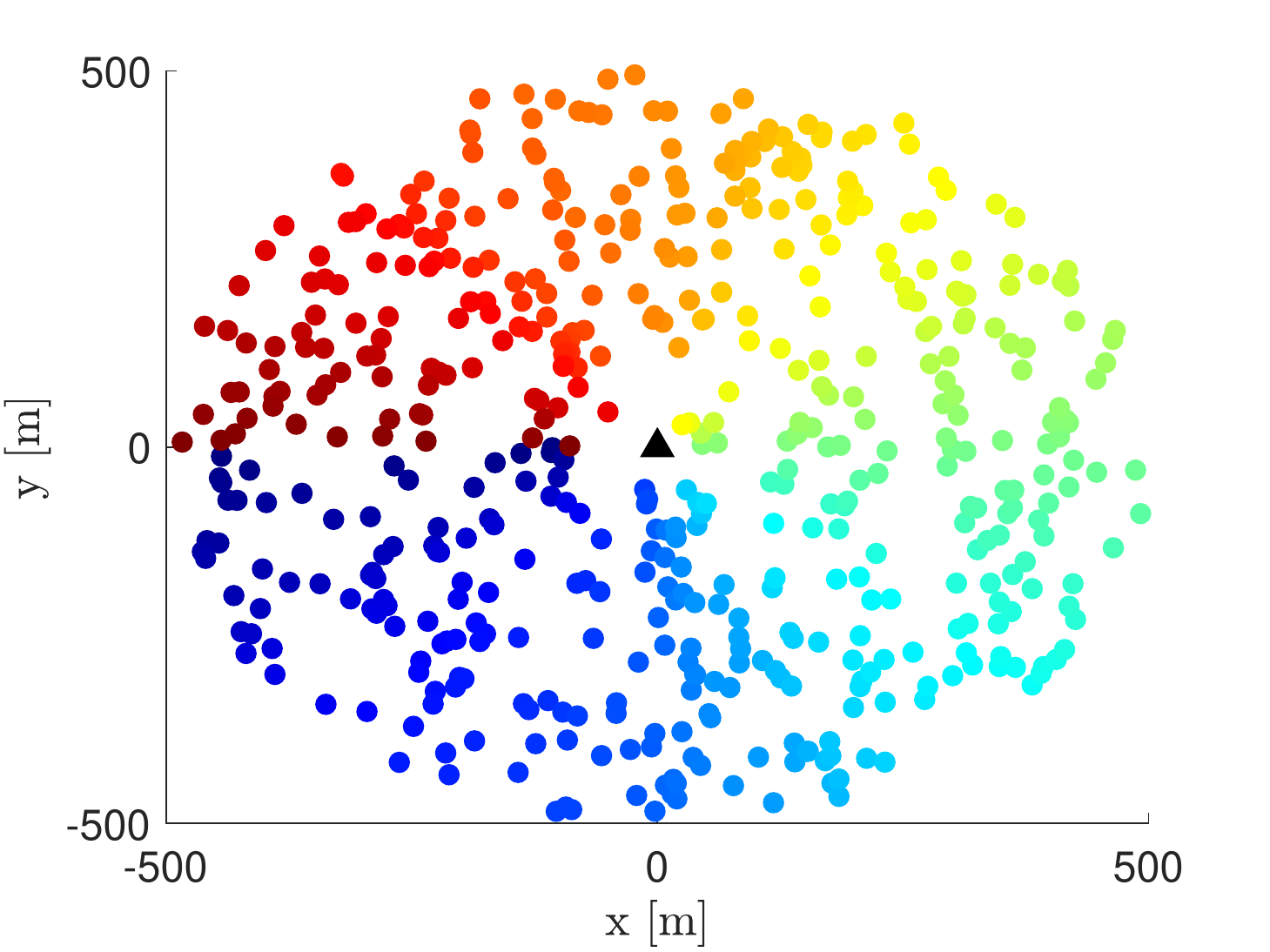}
    \caption{UE deployment.}
    \label{subfig:UEs}
    \end{subfigure}
    \begin{subfigure}{0.45\textwidth}
    \includegraphics[width=\textwidth]{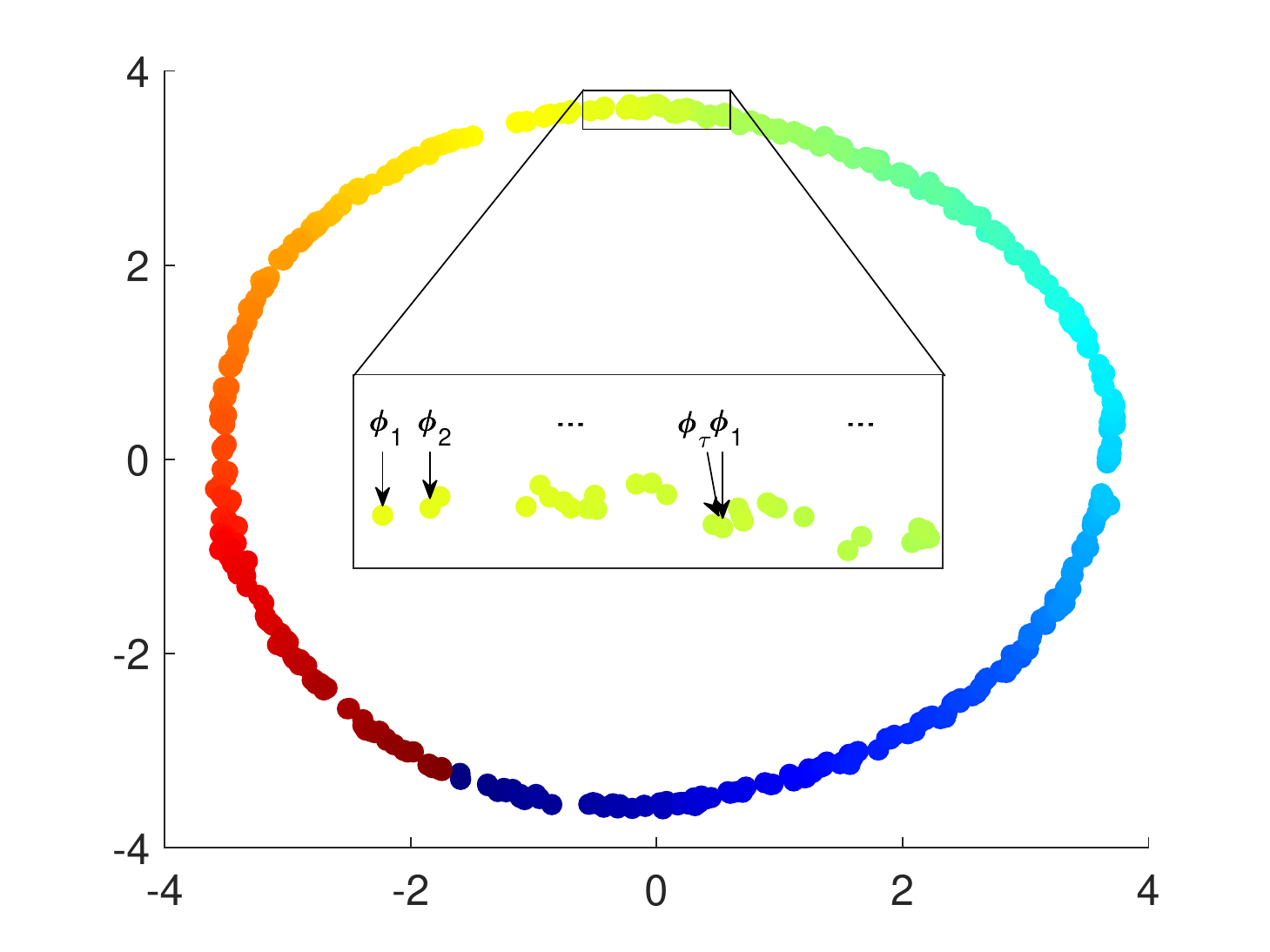}
    \caption{2D CC.}
    \label{subfig:CC1D}
    \end{subfigure}\\
    \begin{subfigure}{0.45\textwidth}
    \includegraphics[width=\textwidth]{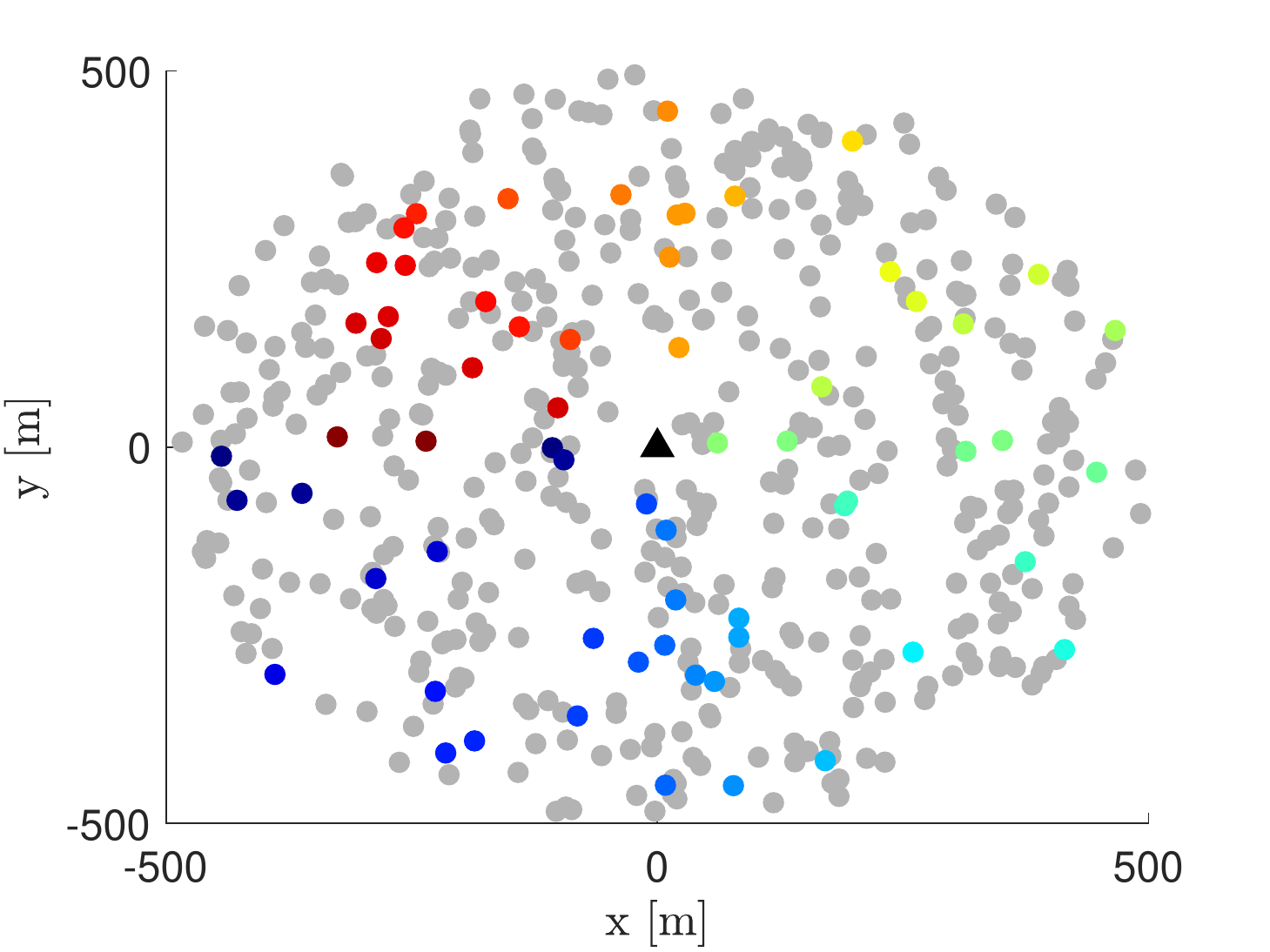}
    \caption{Active UEs.}
    \label{subfig:activeUEs}
    \end{subfigure}
    \begin{subfigure}{0.45\textwidth}
    \includegraphics[width=\textwidth]{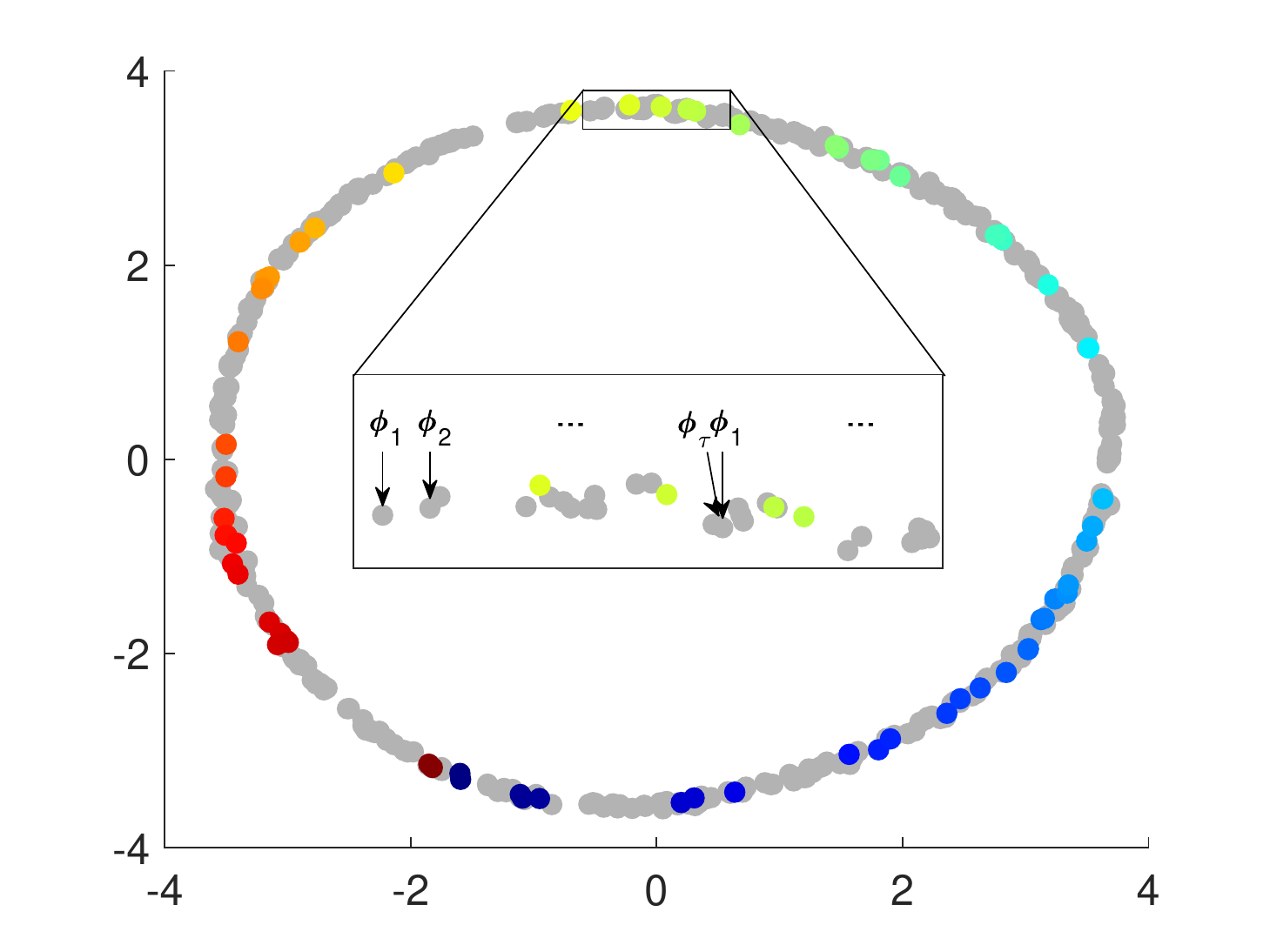}
    \caption{2D CC highlighting active UEs.}
    \label{subfig:CCactive1D}
    \end{subfigure}
    \caption{Illustration of the proposed CC-based pilot allocation algorithm. (a) shows the physical locations of 512 UEs deployed in a cell with $S=3$ ULAs and (b) depicts the pilot assignment for a fraction of the UEs, detailed in the zoomed region. (c) shows the same deployment as in (a), yet highlighting only the set of active UEs and (d) shows the corresponding pilot assignment.}\vspace{-4mm}
    \label{fig:CC-pilotAllocation}
\end{figure*}

\subsubsection{Isomap}
Isomap is a DR technique that aims to preserve the geodesic distance or the curvilinear distance measured over the higher dimensional space between the features~\cite{Tenenbaum2000}. The geodesic distance is computed by constructing a neighborhood graph $\mathrel{G}$, in which every UE is connected with its $\nu$ nearest neighbors, where the integer $\nu$ is a design parameter~\cite{Maaten2009}.

The first step of the Isomap algorithm is to find the $\nu$ closest (neighboring) UEs for each UE in the feature domain, represented by ${\vec{F}\in\mathbb{R}^{N\times N}}$. To find the ${\nu}$ neighbors of UE $n$, we look for the ${\nu+1}$ smallest elements in ${\vec{f}_n}$. Then, we construct a weighted graph $\mathrel{G}$ based on neighbor information. The weights in $\mathrel{G}$ connecting two vertices, i.e., UEs, that are deemed to be among the nearest neighbors correspond to the distances in $\vec{F}$ between these two UEs, whereas the weights of non-nearest neighbors are set to a large number. Note that $\mathrel{G}$ is a directed graph, i.e., UE $j$ may not be in the neighborhood of UE $n$, even if UE $n$ is in the neighborhood of $j$.

After acquiring the graph $\mathrel{G}$, one needs to compute the geodesic distances across $\mathrel{G}$ using a shortest path algorithm, such as Dijkstra~\cite{Dijkstra} or Floyd's~\cite{Floyd} algorithm. The output of the shortest path algorithm is a dissimilarity matrix ${\vec{F}^{\prime}\in\mathbb{R}^{N\times N}}$ which consists of geodesic distances computed along the nearest-neighbor graph.

The last step of Isomap consists of applying classical multidimensional scaling (MDS) to the dissimilarity matrix ${\vec{F}^{\prime}}$~\cite{Tenenbaum2000}. To start with this, we compute the doubly centered dissimilarity matrix ${\vec{K}^{\prime}=-\frac{1}{2}\vec{C}{(\vec{F}^{\prime}\odot\vec{F}^{\prime})}\vec{C}}$. Next, we extract the CC coordinates by taking the eigenvalue decomposition of ${\vec{K}^{\prime}}$. Let ${\lambda_1^{\prime},\lambda_2^{\prime},\ldots,\lambda_{C}^{\prime}}$ be the $C$ largest eigenvalues of ${\vec{K}^{\prime}}$ and ${\vec{u}_1^{\prime},\vec{u}_2^{\prime},\ldots,\vec{u}_{C}^{\prime}}$ the associated eigenvectors. Finally, the CC mapping, when utilizing Isomap as the DR technique, is obtained as
\begin{equation}
    \vec{Z}^{\mathrm{ISO}}_{C}=\big[\sqrt{\lambda_{1}^{\prime}}\vec{u}_1^{\prime},\ldots,\sqrt{\lambda_{C}^{\prime}}\vec{u}_{C}^{\prime}\big]\tran.
    \label{eq:CC-ISO}
\end{equation}

The dimensionality of the final embedding affects the accuracy of the representation of the features in the lower dimensional chart. This is illustrated in Fig.~\ref{fig:CC}. In Fig.~\ref{fig:CC}(a), ${N=40}$ UEs are uniformly distributed in a 1~km$^2$ cell and colored with respect to their AoAs to the BS. Figs.~\ref{fig:CC}(b), (c), and (d) depict the CC mapping obtained for ${C=1}$, ${C=2}$ and ${C=3}$ dimensions, respectively. Here, one can clearly see that the accuracy of the angular domain representation increases as $C$ increases. 

In quantitative terms, the accuracy of the CC embedding is measured by the residual variance between the feature distances in the high-dimensional space, ${\vec{F}}$ in~\eqref{eq:CMD_centralized}, and in the CC domain, ${\vec{D}_{\vec{Z}_C}=[\vec{d}_1,\ldots,\vec{d}_N]\in\mathbb{R}^{N\times N}}$. The $(n,j)\text{-th}$ entry of the distance matrix ${\vec{D}_{\vec{Z}_C}}$ is computed as the ${\ell_2\text{-norm}}$ of the difference between the CC points ${\vec{z}_n}$ and ${\vec{z}_j}$, i.e., ${\left[\vec{D}_{\vec{Z}_C}\right]_{n,j}=\norm{\vec{z}_n-\vec{z}_j}}_2$.
Therefore, the residual variance of the CC mapping is given by
\begin{equation}
\label{eq:residualVariance}
    r\left(\vec{F},\vec{D}_{\vec{Z}_C}\right)=1-\frac{\sum_{n=1}^{N}(\vec{f}_n-\overbar{\vec{f}_n})\tran(\vec{d}_n-\overbar{\vec{d}_n})}{\sqrt{\mathrm{Var}(\vec{F})\mathrm{Var}(\vec{D}_{\vec{Z}_C})}},
\end{equation}
where ${\mathrm{Var}(\vec{F})=\sum_{n=1}^{N}(\vec{f}_n-\overbar{\vec{f}_n})\tran(\vec{f}_n-\overbar{\vec{f}_n})}$ 
and ${\mathrm{Var}(\vec{D}_{\vec{Z}_C})=\sum_{n=1}^{N}(\vec{d}_n-\overbar{\vec{d}_n})\tran(\vec{d}_n-\overbar{\vec{d}_n})}$. 
Note that the second term in~\eqref{eq:residualVariance} is the Pearson’s correlation coefficient for the vectorized distance matrices ${\vec{F}}$ and ${\vec{D}_{\vec{Z}_C}}$~\cite[p. 157]{Milton2003}.
In Section~\ref{sec:pilotStrategy}, the residual variance is utilized to find a good compromise between the number of dimensions of CC and the accuracy of the features' representation in the CC, which impacts the performance of the proposed pilot assignment algorithm.

\section{PILOT REUSE ALGORITHM}
\label{sec:pilotStrategy}
To minimize the total amount of pilot contamination as formulated in~\eqref{eq:assignment-problem}, we propose a pilot reuse strategy that exploits the interference map generated through CC, as presented in Section~\ref{sec:CC}, to assign orthogonal pilot sequences to UEs inducing potentially strong mutual interference. The pilot sequences are allocated to all UEs regardless of their activity pattern. Therefore, reassignment is needed only if the network setup or the UEs' relative positions vary considerably. 
Because the interference pattern depends on the UEs' relative positions, if the latter changes, the pilot sequences must be reassigned to accommodate the new interference pattern. In practice, the relative positions of UEs are related to their physical positions, namely their angular positions. Therefore, if the UE mobility is high, the pilot allocation algorithm has to be re-executed more frequently.
To handle cases where few devices join or disconnect from the network, out-of-sample CC~\cite{Ponnada2019} along with a pilot allocation margin could be implemented to avoid re-creating the whole UE-interference map and reassigning all the pilot sequences. 

Furthermore, we propose an adaptive CC to recover the interference map with a flexible number of dimensions. 
This additional flexibility in the UE-interference map increases the robustness of the method to changes in the UEs' density by preserving the meaningful structures embedded in the features while discarding unnecessary dimensions. That improves the CC interference map representation, in particular, for low-density (sparse) deployments.

\subsection{CC-AIDED PILOT ASSIGNMENT}
The proposed CC-aided pilot reuse scheme for single-cell mMIMO deployments is divided into two main phases: 1) the acquisition of the UEs' interference map, and 2) the low-complexity nearest neighbor pilot assignment algorithm.\vspace{-4mm}

\subsubsection{Adaptive CC Interference Map}
To generate the interference map with adaptive dimensions, we apply Isomap with the same principles as described in {Section~\ref{sec:CC}.\ref{subsec:DR}}. However, instead of fixing $C$ from the beginning, we propose to start with a small number of dimensions and increase it to meet the two mapping accuracy criteria described below. In practice, we start with $C=1$ and increase the dimensionality of the resulting embedding only if the residual variance of CC given by~\eqref{eq:residualVariance} is higher than a threshold $\epsilon$ or if it varies less than $\xi$ from increasing $C$, i.e., ${r(\vec{F},\vec{Z}_C)>\epsilon}$ or ${|r(\vec{F},\vec{Z}_C)-r(\vec{F},\vec{Z}_{C-1})|>\xi}$. The thresholds $\epsilon$ and $\xi$ are design parameters to tune the number of dimensions underlying the interference information embedded on $\vec{F}$. To shed some light, based on our empirical observations, ${\epsilon = 0.0001}$ and ${\xi=0.0001}$ present a good compromise between the number of dimensions required and the residual variance to represent the features' embedding. Also, the numerical results will show that the number of dimensions required to have an accurate interference map depends on the density of UEs.\vspace{-4mm}
\subsubsection{Greedy Pilot Allocation}
\begin{algorithm}[t]
\caption{Nearest neighbor pilot allocation}
\label{alg:NNPA}
\SetAlgoNoLine
\DontPrintSemicolon
\SetKwInOut{Input}{Input}\SetKwInOut{Output}{Output}
\SetKwFor{For}{for}{do}{end for}

\Input{1) The set of UEs $\mathcal{N}=\{1,\ldots,N\}$, 2) the index set of orthogonal pilots $\mathcal{T}=\{1,\ldots,\tau\}$, and 3) the suitable features $\vec{z}_n,\, n \in\,\mathcal{N}$.}

\Output{A pilot assignment $\vec{\Psi}=\left[\vec{\psib}_1,\ldots,\vec{\psib}_N\right]\tran$.}

Initialize the set of unassigned UEs $\mathcal{N}^{\mathrm{un}}=\mathcal{N}$, and the set of unassigned pilots $\mathcal{T}^{\mathrm{un}}=\mathcal{T}$.\;

Select a random UE $n$ and initialize the auxiliary variable $n'$ with it, i.e., $n'=n$.\;

Assign $\phib_1$ to user $n$ and update the set of unassigned UEs and pilots, i.e., $\mathcal{N}^{\mathrm{un}}\gets \mathcal{N}^{\mathrm{un}}\setminus \{n\}$ and $\mathcal{T}^{\mathrm{un}}\gets \mathcal{T}^{\mathrm{un}}\setminus \{1\}$.\;

Initialize the auxiliary variable: $p=2$.\;
\While{$\mathcal{N}^{\mathrm{un}}\neq \emptyset$}{
  \If{$\mathcal{T}^{\mathrm{un}}=\emptyset$}{
  Reinitialize: $\mathcal{T}^{\mathrm{un}}=\mathcal{T}$ and $p=1$.\;
   }
   
  Assign pilot $\phib_{p}$ to user $n$, i.e., $\vec{\psib}_n=\phib_p$, that satisfies $n=\underset{n\in \mathcal{N}^{\mathrm{un}}}{\argmin}\Vert \vec{z}_n-\vec{z}_{n'}\Vert^2$.\;
  
  Update the set of unassigned UEs, $\mathcal{N}^{\mathrm{un}}\gets \mathcal{N}^{\mathrm{un}}\setminus \{n\}$, and the set of unassigned pilots, $\mathcal{T}^{\mathrm{un}}\gets \mathcal{T}^{\mathrm{un}}\setminus \{p\}$.\;
  
  Update $n'=n$ and $p=p+1$.\;
 }
 \vspace{-1mm}
 \end{algorithm}
\begin{figure}[t]
    \centering
    \includegraphics[width=0.52\textwidth,trim=3cm 4cm 4cm 4cm,clip]{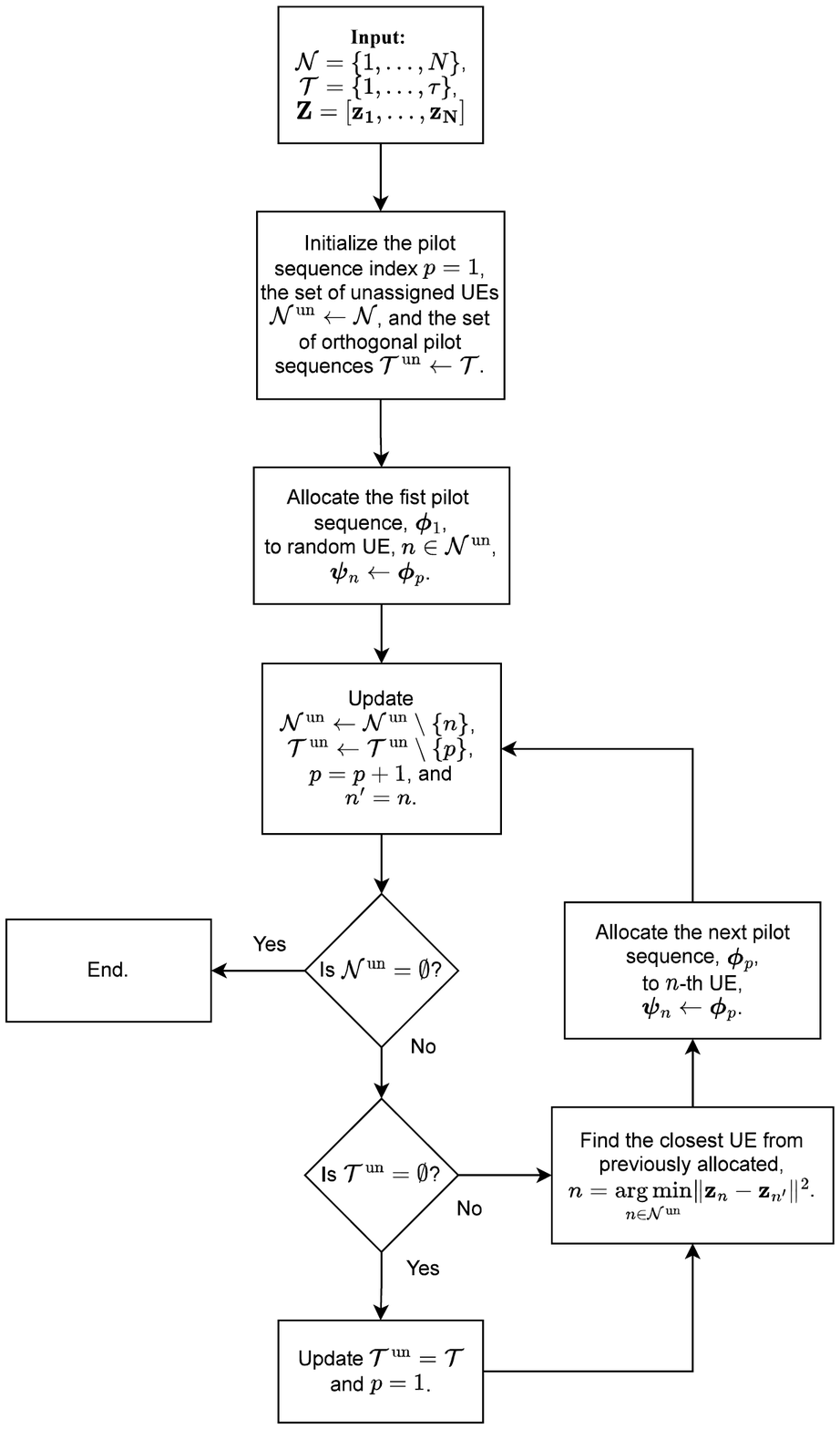}
    \caption{Flow chart of the proposed nearest neighbor pilot allocation algorithm. The algorithm greedily allocates the pilot sequences to maximize the distance between the same pilot sequences in the CC domain.}
    \label{fig:pilotAllocationDiagram}
    \vspace{-0.7cm}
\end{figure}
The problem still remains weighted graph coloring, and thus NP-hard, but with distances computed in reduced dimensions. Intuitively, finding a solution has become easier, as the applied DR mechanism reduces the effect of the bulk of small interferences and enhances the large interferences, thus reducing the probability of falling to a bad local minimum. We apply greedy allocation of orthogonal pilots to the nearest UEs (neighbors) in the CC domain, as described in Algorithm~\ref{alg:NNPA} and depicted on the flow chart in Fig.~\ref{fig:pilotAllocationDiagram}. Having acquired an interference map where the strongly interfering UEs are mapped closer than the weakly ones, the main idea of the proposed pilot assignment algorithm is to allocate the pilot sequences in an ordered way, ${\phib_1,\phib_2,\ldots,\phib_\tau,\phib_1,\phib_2,\ldots}$, aiming at maximizing the distances between the same pilot sequences on the interference map. This is equivalent to maximizing the AoA distances between UEs that have the same pilot sequences. We delve into details of the proposed algorithm in the following. 

To start the pilot allocation, the first orthogonal pilot sequence, ${\phib_1}$, is assigned to a randomly chosen UE. In sequence, the next pilot sequence is assigned to the unassigned UE with the smallest Euclidean distance to the previously allocated user, i.e., ${\phib_2}$ is allocated to the UE $n$ that minimizes ${\norm{\vec{z}_n-\vec{z}_{n'}}^2_2}$, where $n'$ is the index of the previously allocated UE to ${\phib_1}$. This process is repeated until all orthogonal pilot sequences have been allocated. From this point onward, the pilot sequences are to be reused by allocating the first sequence, ${\phib_1}$, to the closest unassigned UE from the precedent allocated user. This process is repeated until all UEs have been assigned a pilot sequence.

Fig.~\ref{fig:CC-pilotAllocation} exemplifies the pilot allocation algorithm procedure for a cell with ${N=512}$ UEs. Fig.~\ref{fig:CC-pilotAllocation}(a) shows the UEs' real positions, which are colored based on their AoA with respect to the BS, and Fig.~\ref{fig:CC-pilotAllocation}(b) shows the respective two-dimensional CC, when Isomap is used as the DR technique. The proposed pilot allocation procedure, as described in {Algorithm ~\ref{alg:NNPA}}, 
is exemplified at the zoomed region in Fig.~\ref{fig:CC-pilotAllocation}(b). Figs.~\ref{fig:CC-pilotAllocation}(c) and (d) illustrate the effect of UEs' activity on the pilot assignment for a random set of active UEs. Comparing Fig.~\ref{fig:CC-pilotAllocation}(b) and Fig.~\ref{fig:CC-pilotAllocation}(d), one can see that the UEs’ activity does not affect the pilot assignment. Moreover, comparing Fig.~\ref{fig:CC-pilotAllocation}(b) with Fig.~\ref{fig:CC}(b), we see that a two-dimensional CC may be a good representation of the angular relationship between UEs when their density in the cell is high. This evidences that the density of UEs in the cell affects the capability of Isomap to accurately map the underlying interference pattern among the UEs into a low-dimensional chart.

\subsection{COMPLEXITY ANALYSIS}
We evaluate the total complexity order of the proposed algorithm through $\mathcal{O}(\cdot)$ notation as: 1) the complexity of generating the adaptive CC interference map through Isomap and 2) the complexity of allocating the pilot sequences using the nearest neighbor pilot algorithm, presented in Algorithm~\ref{alg:NNPA}.

The complexity of generating the CC through Isomap is divided into computing the feature matrix $\vec{F}$ in~\eqref{eq:CMD_centralized}, finding the $\nu$ nearest neighbors, computing the shortest paths across the neighborhood graph $\mathrel{G}$, and performing classical MDS. The cost of computing the CMD distances between all the UE pairs in the $MS$-dimensional space is ${\mathcal{O}(M^3S^3N^2)}$. The complexity to find the $\nu$ nearest neighbors is ${\mathcal{O}(\nu N^2)}$. To compute the shortest paths across the graph $\mathrel{G}$ using Floyd's algorithm (used in the simulations) requires ${\mathcal{O}(N^3)}$. Note that the complexity to compute the shortest paths can be reduced to ${\mathcal{O}(\nu N^2\log(N))}$ using Dijkstra's algorithm~\cite{Silva2003}. 
Lastly, the complexity of computing the doubly centered matrix $\vec{K}^\prime$ and its eigenvalue decomposition is ${\mathcal{O}(N^3)}$~\cite{Tzeng2008}.

The complexity of the pilot allocation algorithm in Fig.~\ref{fig:pilotAllocationDiagram} is given as the cost to compute the pair-wise distances between the $N$ CC points in the $C$-dimensional embedding, which has the complexity ${\mathcal{O}(CN^2)}$, and then finding the closest unassigned UE which has complexity ${\mathcal{O}(N^2)}$.

In summary, the complexity order of running the proposed algorithm is ${\mathcal{O}\left((M^3S^3+\nu+C+1)N^2+2N^3\right)}$. As comparison, the complexity of SGPS~\cite{You2015}, AoA UE grouping~\cite{Li2018}, and CMD-based~\cite{Ribeiro2021} pilot allocation algorithms are also summarized in Table~\ref{tab:complexity}. The proposed method has the same order of complexity as the CMD method, up to a scalar factor, namely ${\nu + C}$. AoA UE grouping has a much lower complexity than the other baseline methods, only ${\mathcal{O}(N^2)}$. For scenarios where the total number $N$ of UEs is much larger than the number of antenna elements $M$ times the number of sectors $S$, the complexity of the proposed method is lower than that of SGPS. Also, assuming that the CC interference map is utilized for other network management purposes, thereby already available, the cost of assigning the pilots is only that of applying the greedy pilot assignment algorithm, which has order of complexity ${\mathcal{O}\left((C+1)N^2\right)}$.

\begin{table}[t]
\centering
\caption{Algorithm complexity analysis.}
\label{tab:complexity}
\begin{tabular}{cc} \toprule
{\bf Scheme} & {\bf Complexity Order} \\ \midrule
AoA UE grouping~\cite{Li2018} & {\small $\mathcal{O}(N^2)$} \\
SGPS~\cite{You2015} & {\small $\mathcal{O}(M^3S^3N^3)$} \\
CMD~\cite{Ribeiro2021}  & {\small $\mathcal{O}((M^3S^3+1)N^2+N^3)$} \\
CC-aided\tablefootnote{When using Isomap as the DR technique.} & {\small $\mathcal{O}((M^3S^3+\nu+C+1)N^2+2N^3)$} \\
\bottomrule
\end{tabular}
\end{table}

\section{SIMULATION RESULTS}
\label{sec:results}

\begin{table}[t!]
\centering
\caption{Simulation parameters.}
\label{tab:parameters}
\begin{tabular}{lc} \toprule
Cell area & $1$~km$^2$  \\
Number of UEs ($N$) & 512 \\
Number of active UEs ($K$) & 64  \\
Number of sectors ($S$) & 3 \\
Number of antenna elements per ULA ($M$) & 64 \\
Carrier frequency ($f_{\mathrm{c}}$) & 6~GHz \\
Wavelength ($\lambda$) & 0.05~m \\
Number of paths ($L$) & 200 \\
Normalized antenna spacing ($\Delta_{\mathrm{r}}$) & 0.5 \\
Angular standard deviation ($\sigma_{\theta}$) & $10^\circ$ \\
Maximum antenna gain ($G_{\mathrm{A_{max}}}$) & $0$~dB\\
Maximum antenna attenuation (${A_{\mathrm{max}}}$) & $30$~dB\\
Half-power beamwidth (${\theta_{\mathrm{3dB}}}$) & $65^\circ$\\
\bottomrule
\end{tabular}
\end{table}

We consider a mMIMO MTC system with ${N=512}$ UEs uniformly distributed within a $1$~km$^2$ cell area, where ${K=64}$ UEs are simultaneously active at any given transmission interval. The BS is equipped with ${S=3}$ ULAs each having ${M=64}$ critically spaced antenna elements (${\Delta_\mathrm{r} = 0.5}$). The propagation channel between each UE and the BS is modelled as the superposition of ${L = 200}$ paths with a fixed angular standard deviation ${\sigma_\theta=10^{\circ}}$, unless otherwise specified. The maximum antenna gain in~\eqref{eq:AntennaGain} is ${G_{\mathrm{A_{max}}}=0}$~dB, the maximum attenuation is ${A_{\mathrm{max}}=30}$~dB, and the half-power beamwidth is  ${\theta_{\mathrm{3dB}}=65^\circ}$~\cite{TR36873}. Isomap DR is used. The residual variance thresholds are ${\epsilon=0.0001}$ and ${\xi=0.0001}$. Binary phase shift keying (BPSK) is used for the pilots and quadrature phase shift keying (QPSK) for the data transmission. The main simulation parameters are summarized in Table~\ref{tab:parameters}.

Channel estimation accuracy is evaluated in terms of the normalized MSE
\begin{equation}
    \text{NMSE CE}=\frac{\mathbb{E}\left[\fronorm{\vec{\hat{H}}}{\vec{H}}\right]}{\mathbb{E}\big[\norm{\vec{H}}^2_F\big]}.
\end{equation}
The expected value is estimated through Monte Carlo simulation so that the NMSE is approximated by normalized average square error, but we use the acronym NMSE for clarity in the sequel. 
We adopt the robust MMSE receiver derived in~\cite[Eq.~(27)]{You2015}, which considers that only channel estimates are known by the BS and, thus, takes into account the channel estimation error. This receiver combining structure is known for minimizing the MSE for symbol detection (MSE SD) and maximizing the instantaneous SINR simultaneously~\cite{You2015,Bjornson2017}. Therefore, the performance of the proposed method is also evaluated in terms of MSE SD, and average achievable uplink rate per UE
\begin{equation}
\label{eq:rate_k}
r_k=\mathbb{E}\left[\log_2{\left(1+\gamma_k\right)}\right],
\end{equation}
where the UE-specific SINR $\gamma_k$, when employing the robust beamforming, is given by
\begin{equation}
    \gamma_k=\vec{\hat{h}}_k\herm \left(\sum_{j\neq k}{\vec{\hat{h}}_j\vec{\hat{h}}_j\herm+\sum_{p=1}^{K}{\vec{R}_{\vec{\tilde{h}}_p}}+\frac{\sigma^2}{p_\mathrm{u}}\vec{I}_M}\right)^{-1} \vec{\hat{h}}_k.
\end{equation}

\begin{figure}[t!]
    \centering
    \includegraphics[width=0.49\textwidth,trim=1mm 0mm 10mm 3mm,clip]{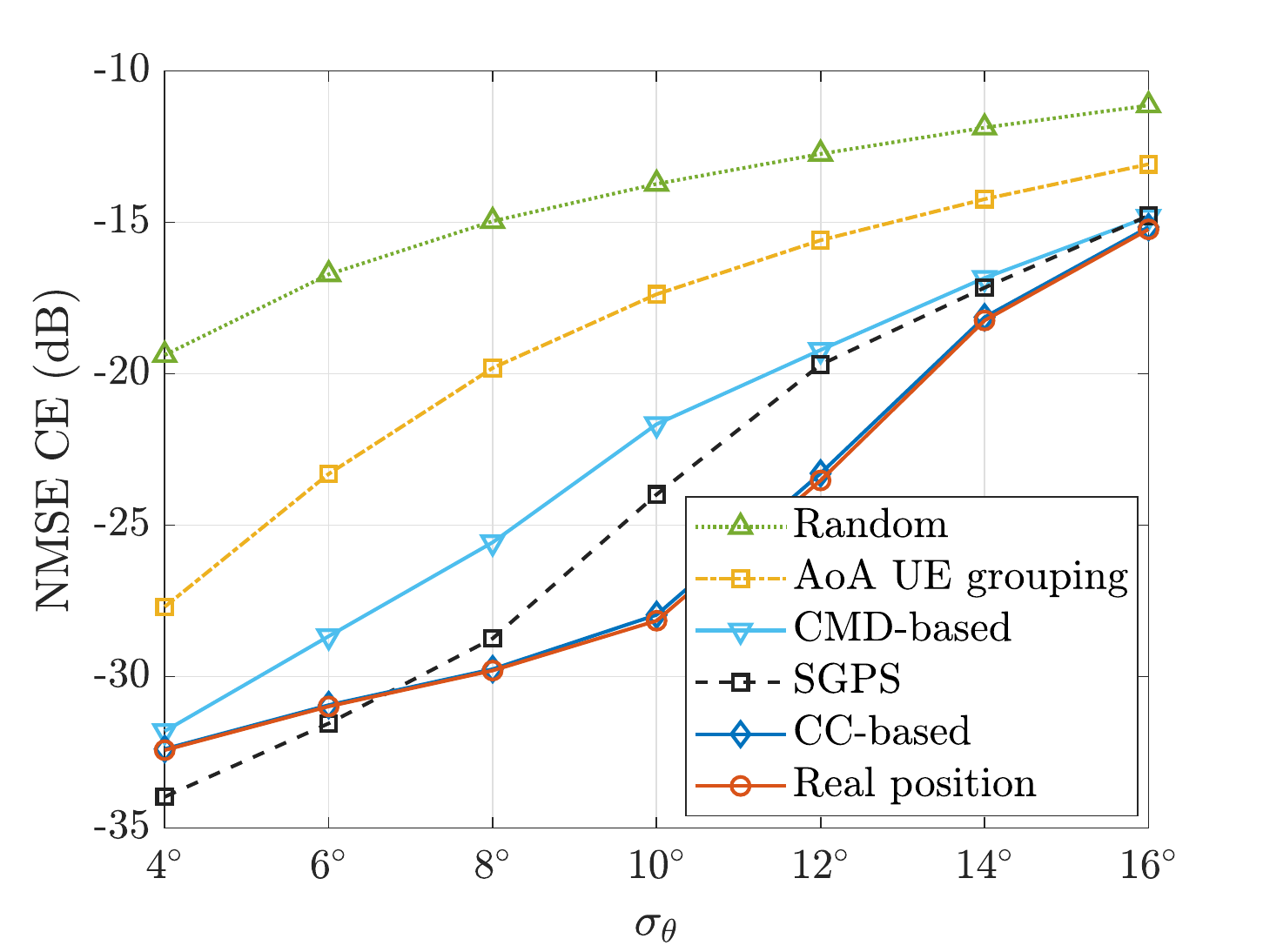}
    \caption{Normalized mean square error for channel estimation versus the angular standard deviation. More directive channels, i.e., narrower angular spread captured by ${\sigma_{\theta}}$, give better channel estimates because the pilots are less likely to interfere with each other.
    }
    \label{fig:NASE-ASD-5-15}\vspace{-8pt}
\end{figure}
Five baseline pilot assignment methods are considered:
\begin{itemize}
    \item ``Random'': This method assigns the orthogonal pilot sequences uniformly at random while respecting the pilot reuse factor, i.e., the same pilot sequence is reused the minimum required times when allocating one sequence to each UE.
    
    \item ``SGPS'': The statistical greedy pilot scheduling (SGPS) developed in~\cite{You2015}, which relies on the knowledge of the channel covariance matrices and mitigates the pilot contamination by grouping the UEs with high spatial correlation and assigns them orthogonal pilot sequences.
    
    \item ``AoA UE grouping'': A multi-cell pilot assignment method proposed in~\cite{Li2018} adapted to the single-cell scenario. This method relies on the exact UEs’ positions to group the UEs based on their AoA. The orthogonal pilot sequences are randomly allocated within each group.
    
    \item ``Real position'': This method assumes knowledge of the exact  physical locations of UEs to compute the angular distances and then applies our nearest neighbor pilot allocation algorithm presented in Algorithm~\ref{alg:NNPA}. Note that this method acts as a lower bound to the proposed one, since it utilizes the real positions of the UEs instead of the relative ones provided by CC.
    
    \item ``CMD'': The CMD-aided pilot assignment method employs the nearest neighbor pilot allocation algorithm (Algorithm~\ref{alg:NNPA}) directly in the feature matrix, $\vec{F}$ in~\eqref{eq:CMD_centralized}, without using any DR technique. This baseline demonstrates the benefit of applying a DR technique in the proposed features.
\end{itemize}
\begin{figure}[t]
\centering
\begin{subfigure}{0.49\textwidth}
    \includegraphics[width=\textwidth,trim=0mm 1mm 10mm 3mm,clip]{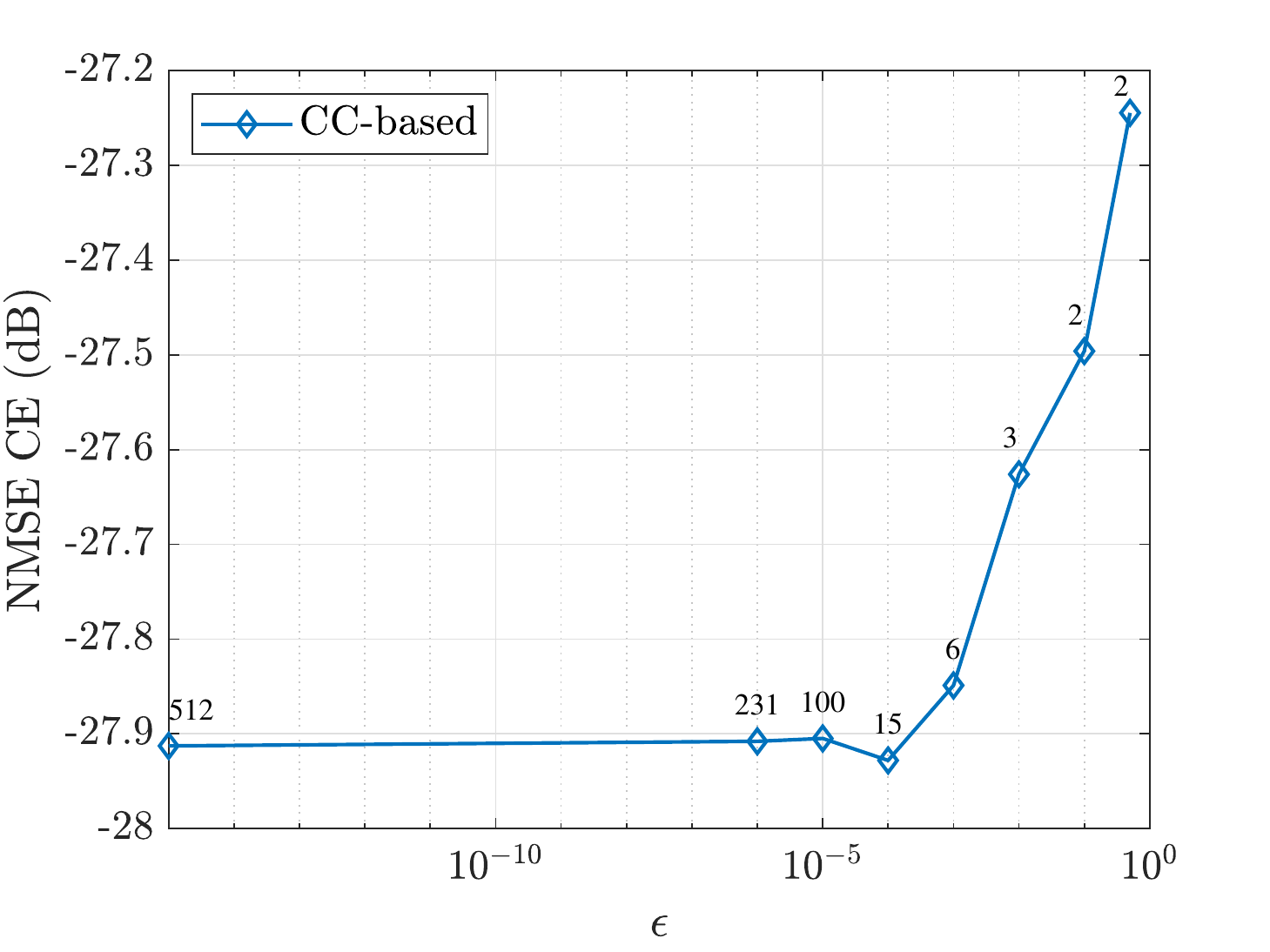}
    \caption{Average NMSE for channel estimation versus $\epsilon$.}
    \label{}
\end{subfigure}
\begin{subfigure}{0.49\textwidth}
    \includegraphics[width=\textwidth,trim=0mm 1mm 10mm 1mm,clip]{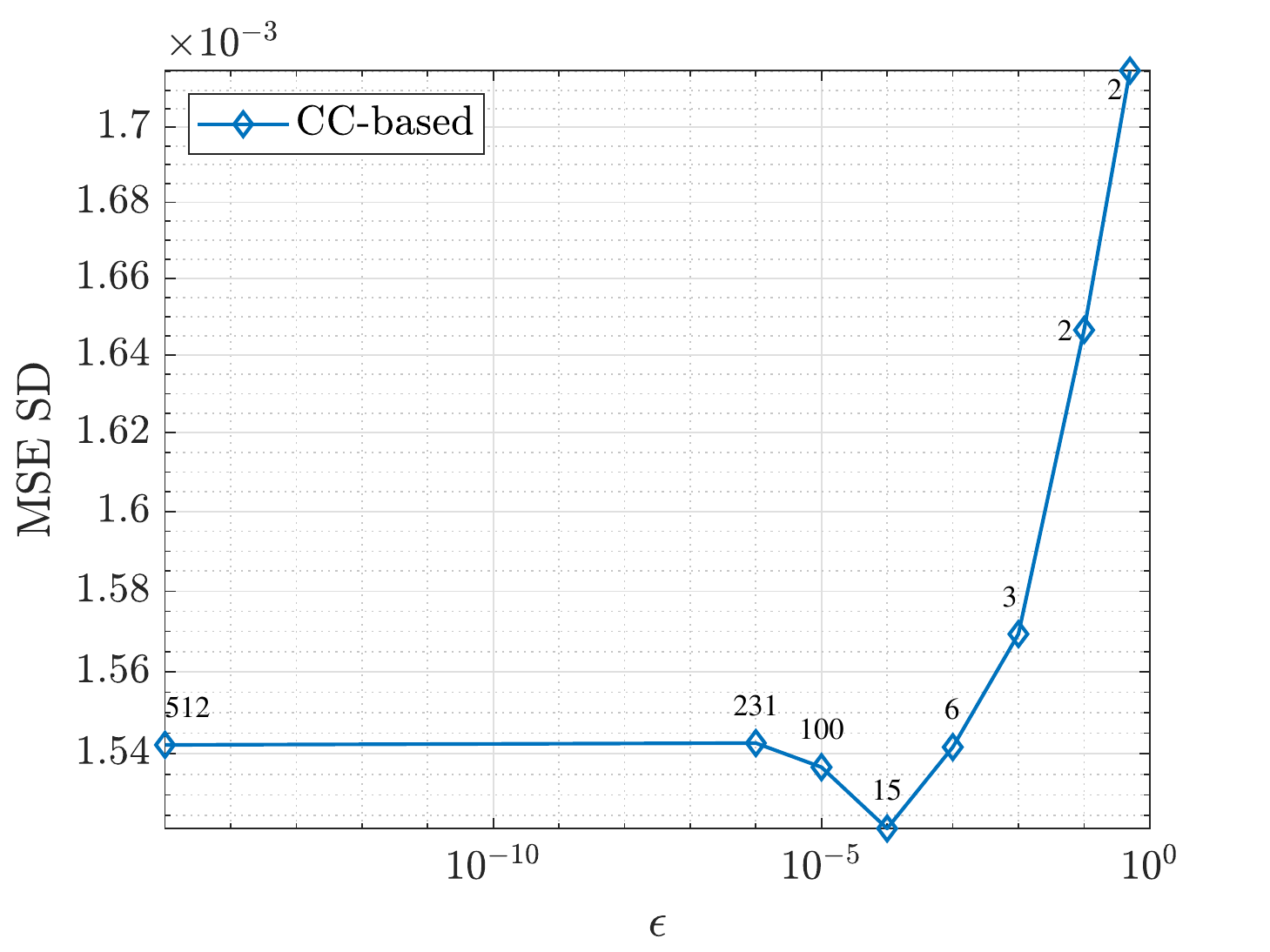}
    \caption{Average MSE for symbol detection versus $\epsilon$.}
    \label{}
\end{subfigure}
\caption{NMSE CE and MSE SD versus threshold parameter $\epsilon$ that controls the dimensionality of CC. The average CC dimension is shown for each value of $\epsilon$ on the top of the curve.}
\label{fig:epsilon}\vspace{-10pt}
\end{figure}

\begin{figure*}[t]
\centering
\begin{subfigure}{0.32\textwidth}
    \includegraphics[width=\textwidth,trim=1mm 1mm 10mm 6mm,clip]{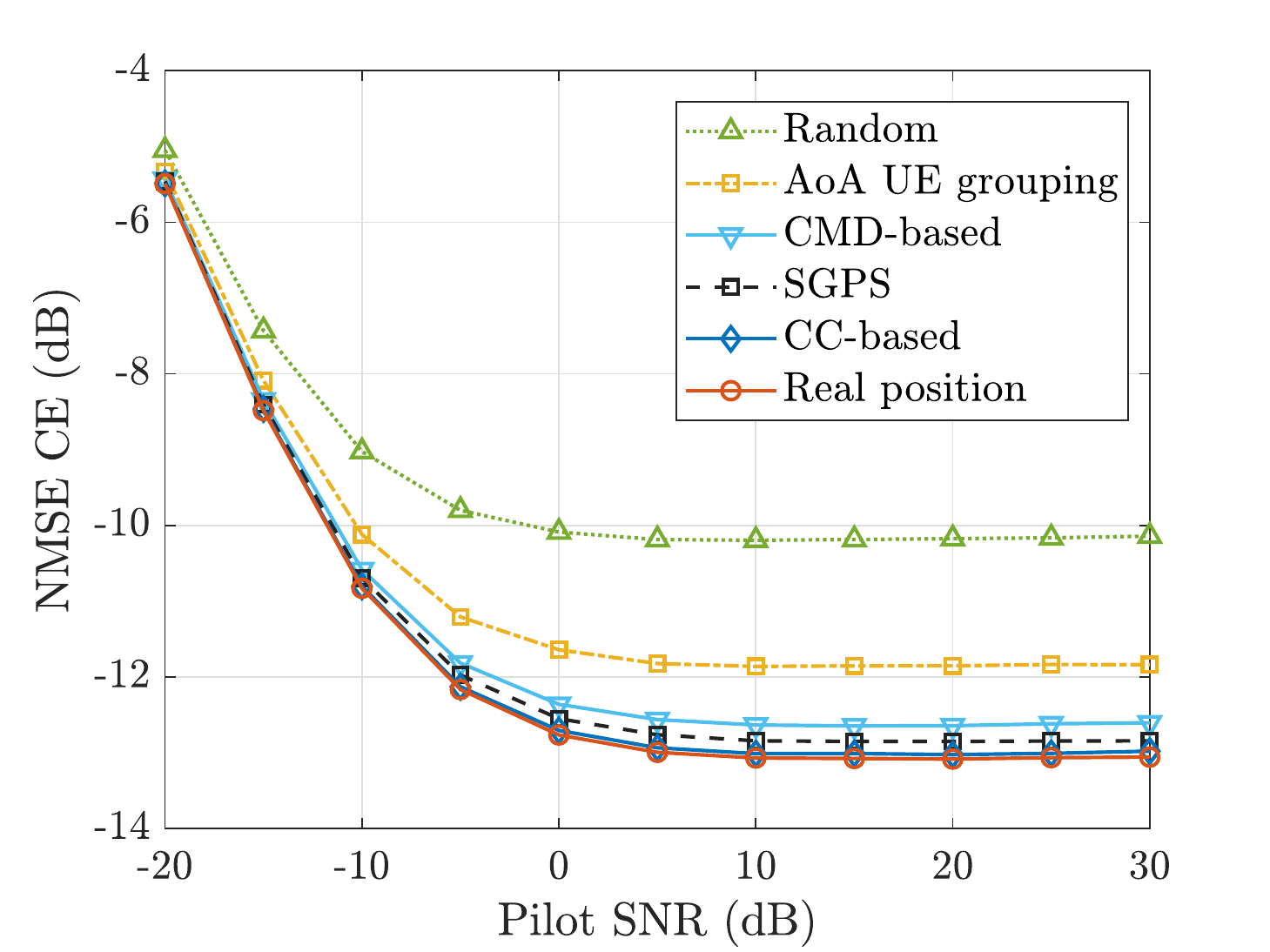}
    \caption{}
\end{subfigure}
\begin{subfigure}{0.32\textwidth}
    \includegraphics[width=\textwidth,trim=1mm 1mm 10mm 6mm,clip]{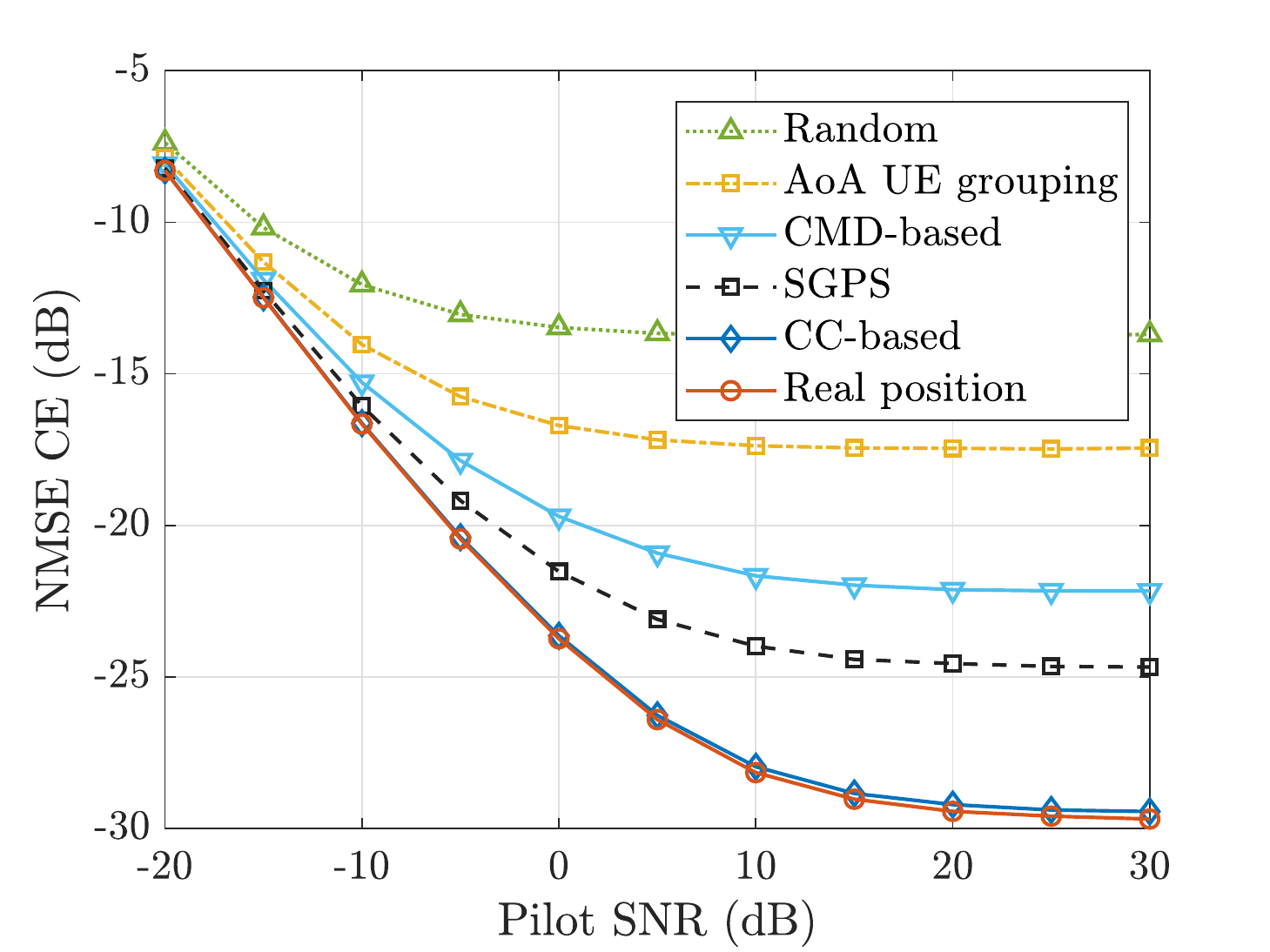}
    \caption{}
\end{subfigure}
\begin{subfigure}{0.32\textwidth}
    \includegraphics[width=\textwidth,trim=1mm 1mm 10mm 6mm,clip]{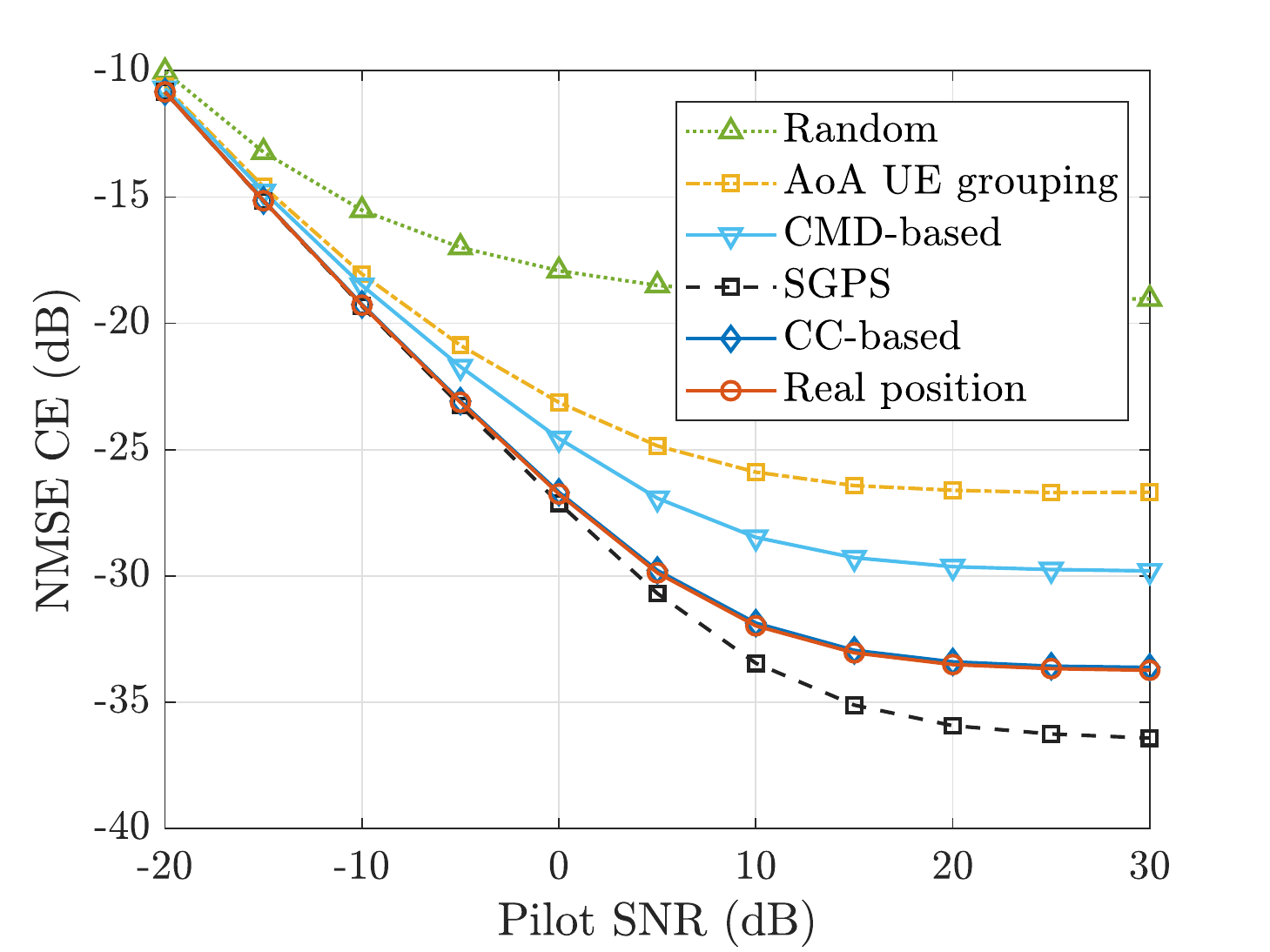}
    \caption{}
\end{subfigure}\\

\begin{subfigure}{0.32\textwidth}
    \includegraphics[width=\textwidth,trim=1mm 1mm 10mm 6mm,clip]{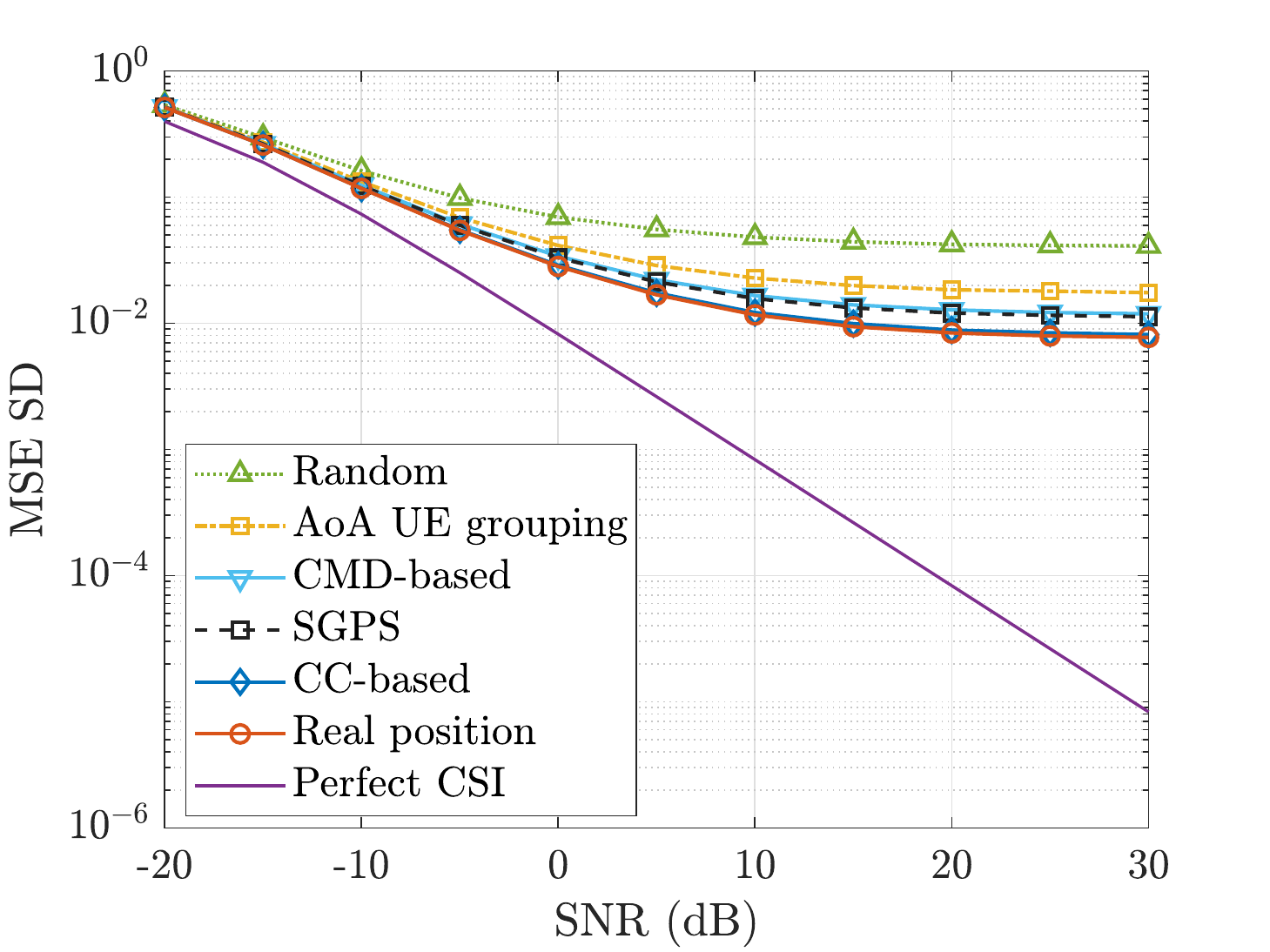}
    \caption{}
\end{subfigure}
\begin{subfigure}{0.32\textwidth}
    \includegraphics[width=\textwidth,trim=1mm 1mm 10mm 6mm,clip]{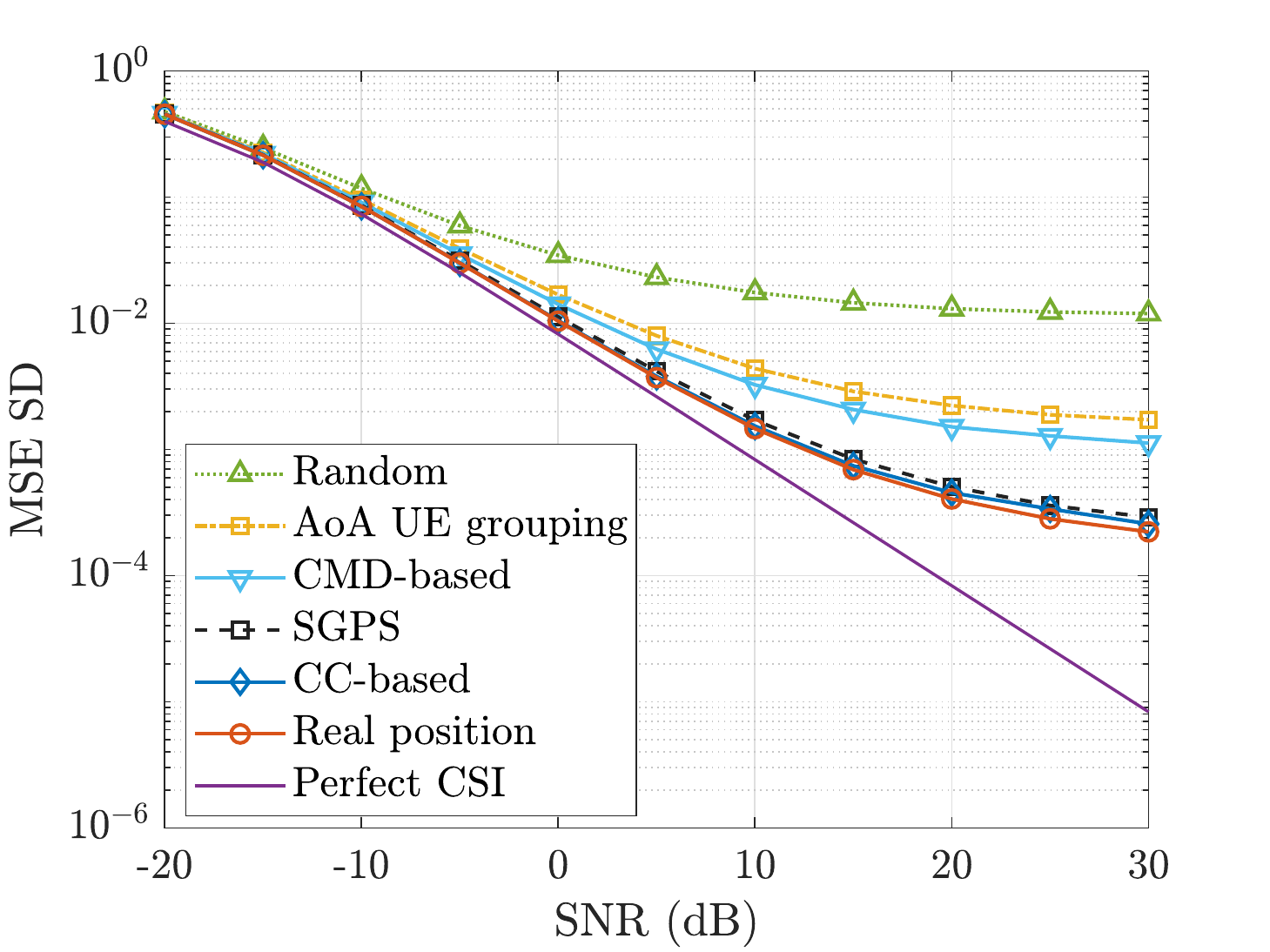}
    \caption{}
\end{subfigure}
\begin{subfigure}{0.32\textwidth}
    \includegraphics[width=\textwidth,trim=1mm 1mm 10mm 6mm,clip]{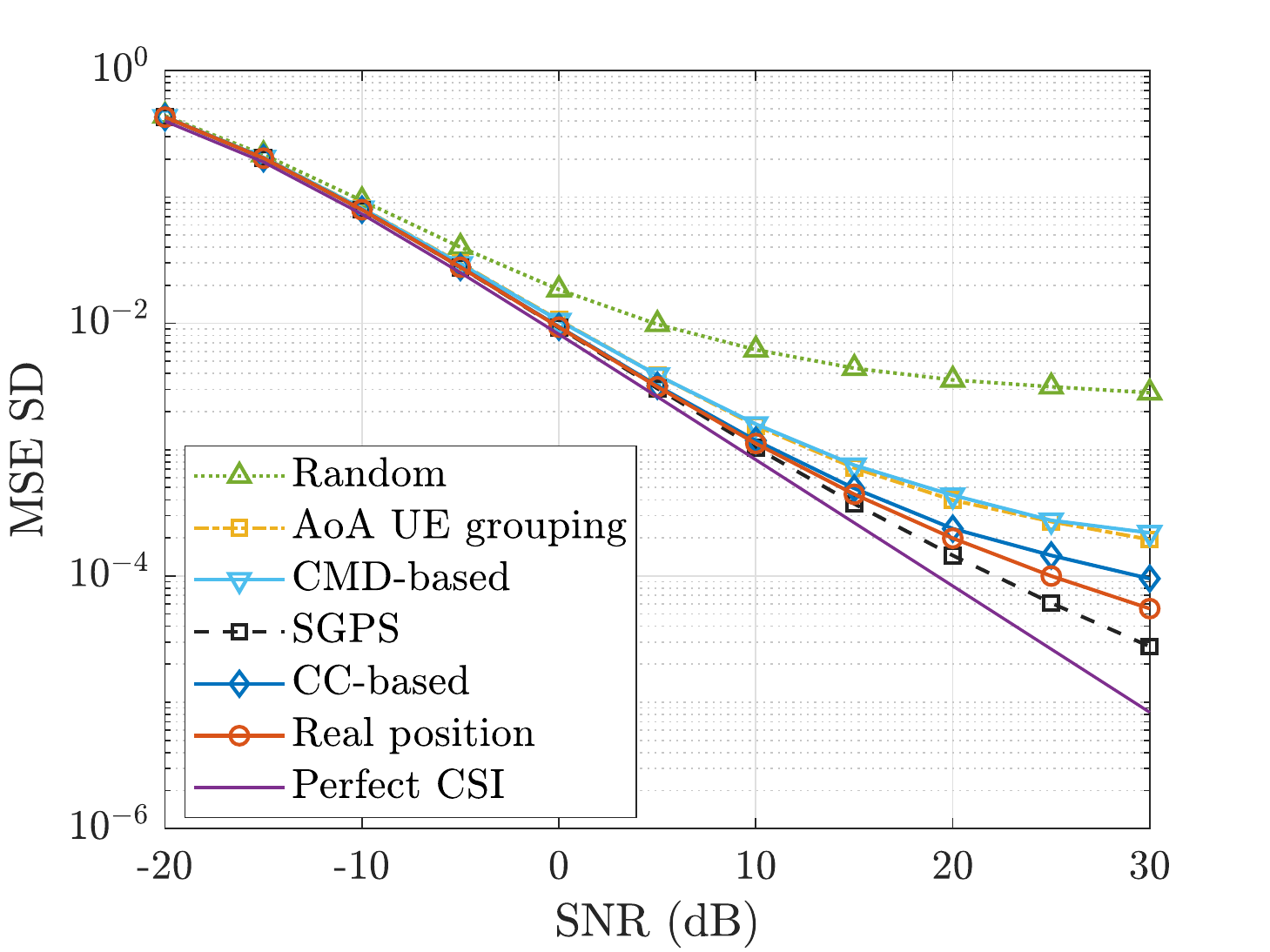}
    \caption{}
\end{subfigure}
\caption{NMSE CE and MSE SD versus the SNR for distinct pilot reuse factors, ${N/\tau=\{16,8,4\}}$, respectively, and for different pilot assignment algorithms. (a) and (d) ${\tau=32}$, (b) and (e) ${\tau=64}$, and (c) and (f) ${\tau=128}$.}
\label{fig:tau-ser}
\end{figure*}

$\text{Fig.\ \ref{fig:NASE-ASD-5-15}}$ presents the performance of the proposed method in terms of NMSE CE for a fixed SNR value of $10$~dB and the angular standard deviation ${\sigma_{\theta}}$ ranging from $4$ to $16$ degrees.  As expected, a smaller angular standard deviation yields a lower channel estimation error. This occurs because the more directive the UEs' channels are, the lower is the interference one UE causes to another. 
Here, we can see that the proposed method works 
well for a moderate and large angular spread of channels, beating all benchmark methods for ${\sigma_{\theta}\geq 8^{\circ}}$, but loses to SGPS algorithm when ${\sigma_{\theta}}$ becomes very narrow. Nonetheless, for such a small angular spread, the performance of the proposed method is as good as the real position benchmark.
As reported in~\cite{Pedersen2000}, reasonable values for the angular standard deviation are between $6^\circ$ and ${21^\circ}$. 
Therefore, we opted for presenting all the subsequent results for ${\sigma_{\theta}=10^{\circ}}$. 
Although the CMD-based method greatly improves the performance as compared to the random pilot allocation and AoA UE grouping, it underperforms compared to the proposed method for all considered angular spreads of channels, highlighting the advantages of preprocessing the features by applying a DR technique.

Fig.~\ref{fig:epsilon} shows how the dimensionality of the CC affects the performance of the proposed method in terms of NMSE CE and MSE SD for ${\sigma_{\theta}=10^\circ}$ and pilot length ${\tau=64}$. The behaviors of the NMSE CE and MSE SD are shown in Fig.~\ref{fig:epsilon}(a) and Fig.~\ref{fig:epsilon}(b), respectively, as a function of the threshold parameter $\epsilon$ for the residual variance in~\eqref{eq:residualVariance}. Also, the average dimensionality of the interference map generated by CC is displayed on the top of the curve. In Fig.~\ref{fig:epsilon}(a), we can see that the performance of the pilot allocation is quite robust to changes in $\epsilon$, i.e., large changes on $\epsilon$ have small effects on the NMSE CE. Therefore, one can reduce the dimensionality of the interference map to decrease computational complexity and storage requirements without losing much in performance. 
In Fig.~\ref{fig:epsilon}, $\xi$ assumes the same value as $\epsilon$ except for $\epsilon=0.1$ and $\epsilon=0.5$ where it takes the values $0.01$ and $0.1$, respectively. Here, one can see that both the NMSE CE and MSE SD reach a point of diminishing returns by reducing $\epsilon$ at ${\epsilon=10^{-4}}$. Therefore, we use ${\epsilon=10^{-4}}$ and $\xi=10^{-4}$ in the subsequent simulations.

Fig.~\ref{fig:tau-ser} shows the NMSE CE and MSE SD as a function of the SNR for distinct pilot lengths, namely ${\tau=\{32,64,128}\}$. One can see that the pilot allocation strategy has a stronger impact at high SNR regimes, where the interference from UEs reusing the same pilot sequence becomes stronger. Note that for large pilot reuse factors, i.e., when the same pilot sequence is reused several times, the gap between the NMSE CE of the methods decreases, like shown in Fig.~\ref{fig:tau-ser}(a). This effect is explained by the exhaustion of the capability of reducing interference, to the extent that even the optimal pilot allocation cannot avoid pilot contamination. That happens because for cases with high reuse of the pilot sequences, the likelihood of UEs with overlapping AoA intervals sharing the same pilot sequence increases. Furthermore, a lower bound for the MSE SD is also presented for benchmark, which is achieved considering perfect CSI knowledge in the LMMSE receive combining. It can be seen that the proposed CC-based method approaches the real position baseline while beating all other methods for moderate and high pilot reuse factors, ${N/\tau=8}$ and $16$, respectively. On the other hand, for a small pilot reuse factor ${N/\tau=4}$, SGPS outperforms the proposed method.

\begin{figure}[t]
    \centering
    \includegraphics[width=0.45\textwidth,trim=1mm 0mm 10mm 3mm,clip]{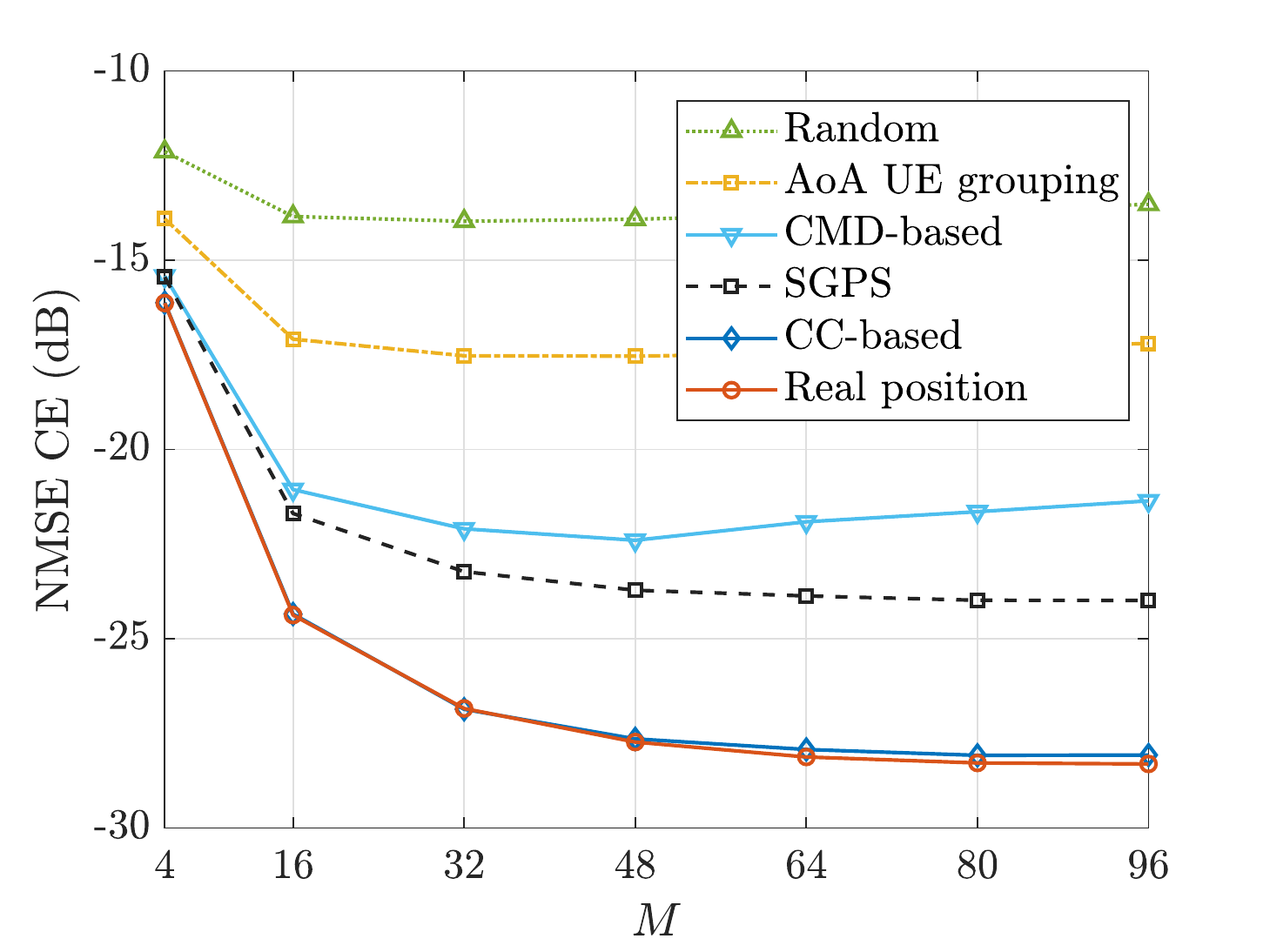}
    \caption{NMSE CE versus the number of antennas $M$ for $N=512$, $K=64$, $\tau=64$, SNR$=10$~dB.}
    \label{fig:NMSE-CE-M}
\end{figure}

The effect of increasing the number of antenna elements on the NMSE CE is presented in Fig.~\ref{fig:NMSE-CE-M}. For fixed SNR and pilot length, one can see that adding more antennas at the BS does not provide additional gains after some threshold. This is due to the fact that the interference is associated with the propagation characteristics of the environment, such as ASD. Thus, increasing the resolvability of the signals via a larger antenna array at the BS may not help distinguish between the UEs' signals. 
In fact, it can even be prejudicial depending on the pilot assignment strategy adopted, as for the CMD-based method in this particular setup. 
This happens because for large antenna arrays more samples are needed to get an accurate covariance matrix~\cite{Ledoit2004}. 
By comparing CMD-based method with CC-based method, it is clear that preprocessing the features before feeding them to the proposed pilot allocation algorithm drastically improves the performance.  

\begin{figure}[t]
    \centering
    \includegraphics[width=0.45\textwidth,trim=1mm 0mm 10mm 3mm,clip]{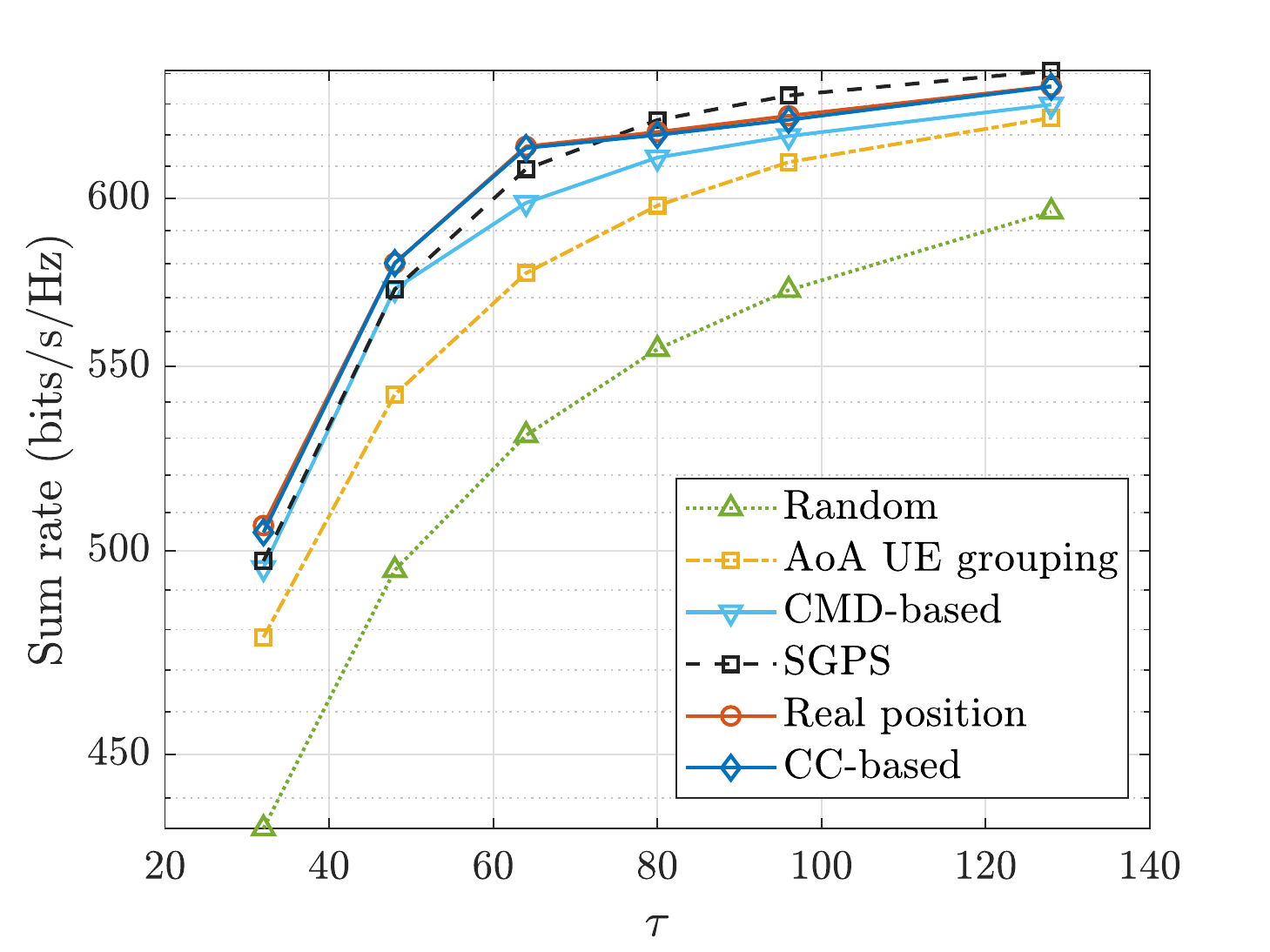}
    \caption{Achievable sum rate for ${K=64}$ active UEs versus pilot length at 10~dB SNR. For ${\sigma_{\theta}=10^{\circ}}$ and different pilot assignment algorithms.}
    \label{fig:SUMRATE-TAU}
\end{figure}

Fig.~\ref{fig:SUMRATE-TAU} shows the uplink achievable rate as a function of the pilot length for ${\text{SNR}=10}$~dB. One can notice that the CC-based method performs very well, approaching the real position based pilot allocation. However, its performance slightly degrades as compared to SGPS for ${\tau>70}$ symbols, i.e., when the pilot reuse factor ${N/\tau}$ decreases below 7. Yet, in this regime, the proposed method performs as well as the real position one.

\begin{figure*}[t]
\centering
\begin{subfigure}{0.32\textwidth}
    \includegraphics[width=\textwidth,trim=1mm 1mm 10mm 6mm,clip]{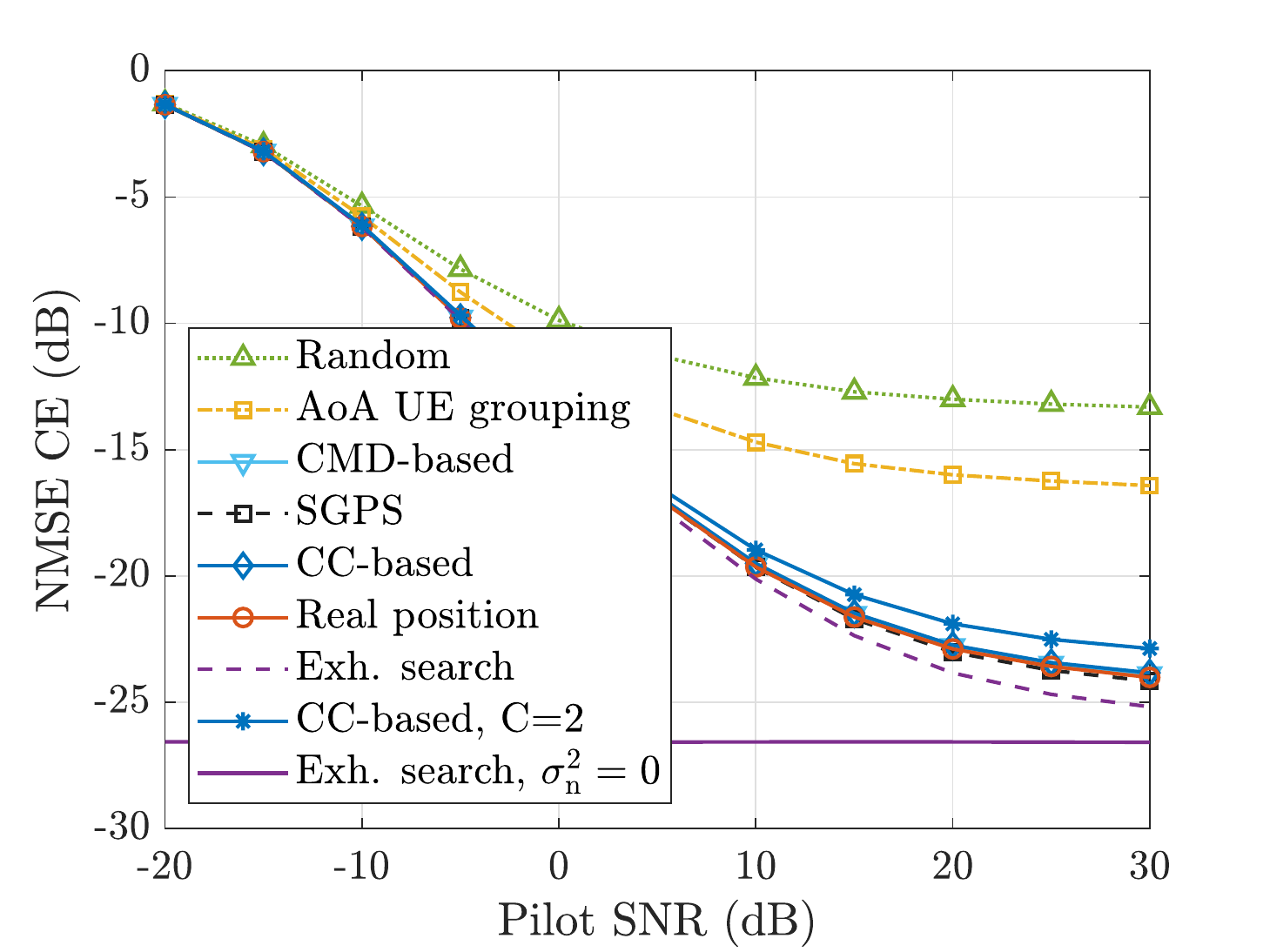}
\end{subfigure}
\begin{subfigure}{0.32\textwidth}
    \includegraphics[width=\textwidth,trim=1mm 1mm 10mm 6mm,clip]{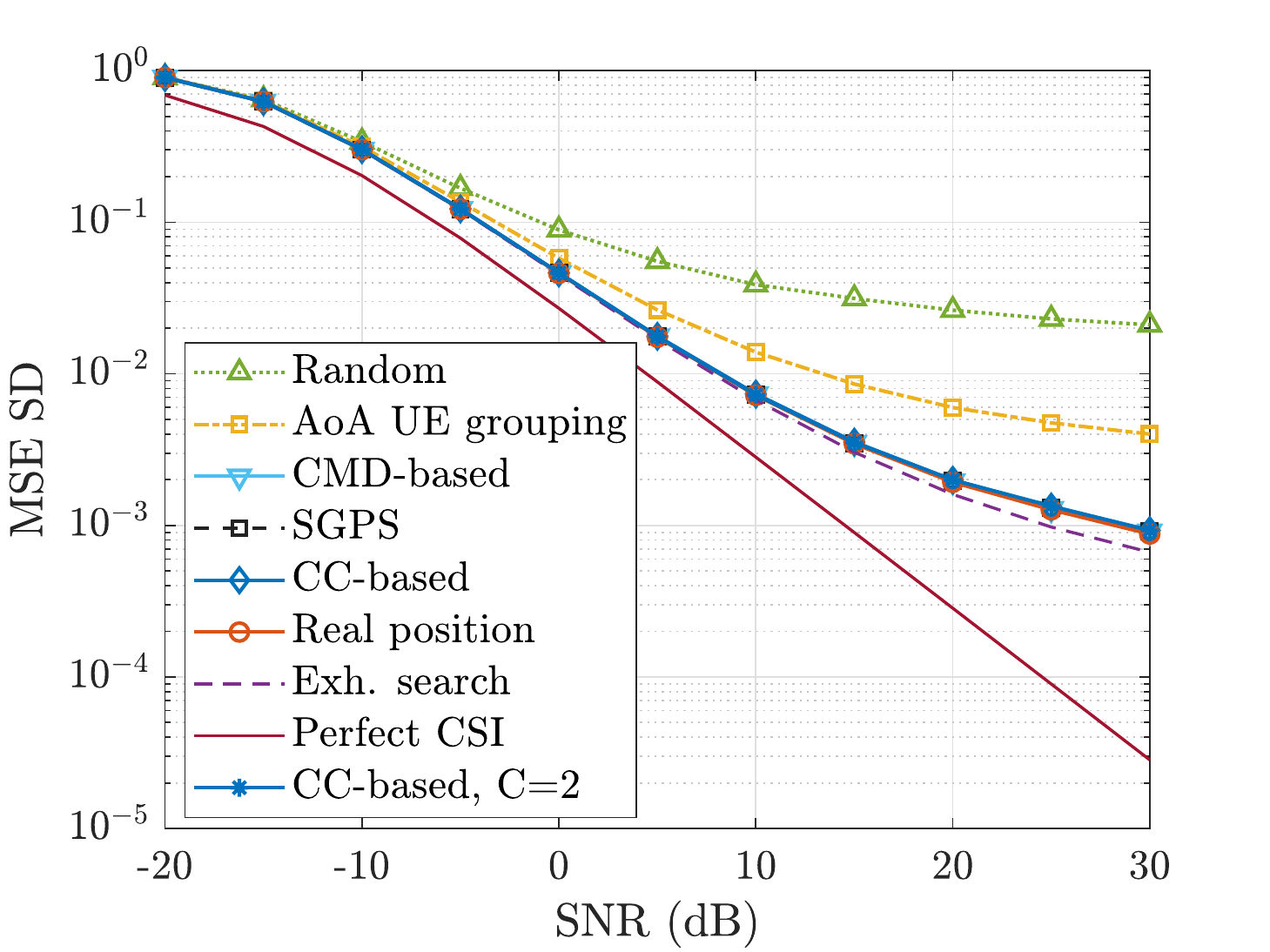}
\end{subfigure}
\begin{subfigure}{0.32\textwidth}
    \includegraphics[width=\textwidth,trim=1mm 1mm 10mm 6mm,clip]{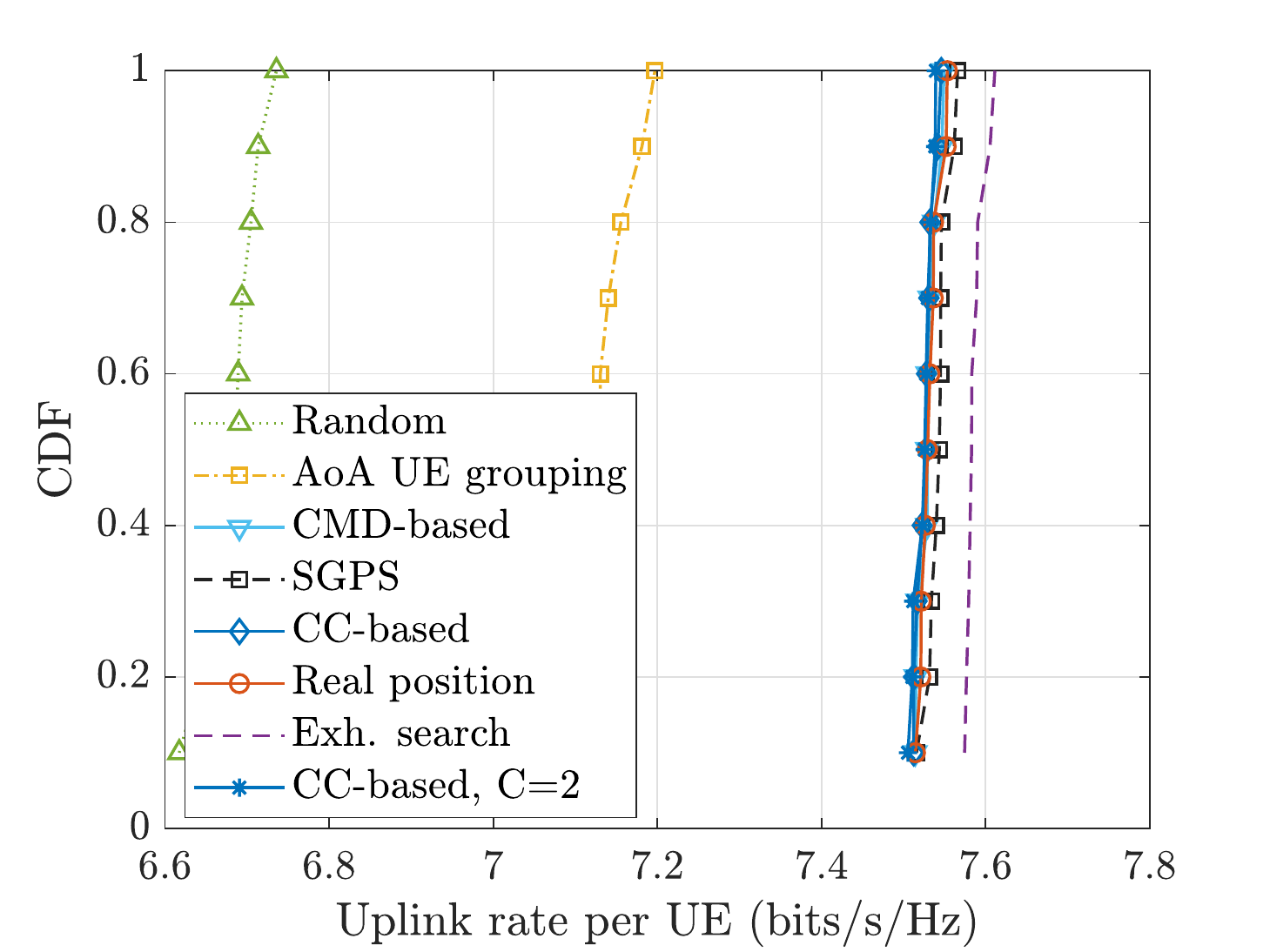}
\end{subfigure}
\caption{The NMSE and MSE-SD as function of SNR and the CDF of the uplink rate per UE, respectively for ${N=10}$, ${K=10}$, ${\tau=4}$, ${\sigma_\theta=10^\circ}$, and ${M=16}$.}\vspace{-4mm}
\label{fig:exhaustiveSearch}
\end{figure*}

The performance of the proposed method is also evaluated against the exhaustive search pilot allocation for a toy scenario in Fig.~\ref{fig:exhaustiveSearch}. Conducting the exhaustive search throughout the orthogonal pilot sequences finds the optimum pilot assignment for~\eqref{eq:assignment-problem}.
In this example, we consider ${N=10}$ UEs, all active, deployed in a cell with ${M=16}$ and ${S=3}$ ULAs. 
To illustrate the effect of the interference on the MSE, we also plot the NMSE CE of the exhaustive search method for the noise-free scenario, i.e., $\sigma_{\mathrm{n}}^2=0$. It is clear from this result that the performance is interference-limited, conditioned on the superposition of UEs' AoA intervals. 
All the methods other than AoA UE grouping and random pilot allocation perform close to the exhaustive search pilot assignment.
The proposed method performs very similarly to the CMD method. That happens, because, for such a small setup with few UEs, more CC dimensions are required to meet the maximum residual variance criterion, ${\epsilon\leq {r(\vec{F},\vec{Z}_C)}}$. That means a low value of $C$ cannot properly reflect the UEs' mutual interference, which is seen here by comparing the performance of the proposed method with the CC-based method with fixed $C$ (CC-based, $C=2$). Thus, CC mapping keeps increasing the number of dimensions to meet the $\xi$ and $\epsilon$ criteria, pushing the performance closer to the one presented by the CMD method.

\section{CONCLUSIONS}
\label{sec:conclusion}
This paper addressed the pilot contamination problem in single-cell multi-sector mMIMO systems. 
We proposed a novel pilot allocation algorithm based on a CC-aided interference map of UEs to cope with the pilot contamination caused by pilot reuse in mMIMO systems. The pilot sequences are assigned so that the angular distances between the UEs that share the same pilot sequence are maximized, which aims at minimizing the interference between UEs. The proposed CC-based pilot allocation algorithm showed significant improvements in terms of channel estimation error, symbol detection error, and achievable uplink rate when compared to existing pilot allocation schemes. In a toy example with few UEs, the performance of the proposed method gets close to the optimal exhaustive search pilot allocation and approaches the real position benchmark for large pilot reuse ratios.

Several avenues for further research remain. The extension of the schemes to the multi-cell setup with cooperative multi-point processing or a cell-free architecture would be of interest. The mMTC literature for grant-free access has also other solutions (such as the use of non-orthogonal pilot sequences) to solve the pilot collision problem. Sequence-design-based solutions for joint channel estimation and UE activity detection with CC could be a fruitful approach for system design.

\bibliographystyle{IEEEtran}

\begin{thebibliography}{10}
\providecommand{\url}[1]{#1}
\csname url@samestyle\endcsname
\providecommand{\newblock}{\relax}
\providecommand{\bibinfo}[2]{#2}
\providecommand{\BIBentrySTDinterwordspacing}{\spaceskip=0pt\relax}
\providecommand{\BIBentryALTinterwordstretchfactor}{4}
\providecommand{\BIBentryALTinterwordspacing}{\spaceskip=\fontdimen2\font plus
\BIBentryALTinterwordstretchfactor\fontdimen3\font minus
  \fontdimen4\font\relax}
\providecommand{\BIBforeignlanguage}[2]{{%
\expandafter\ifx\csname l@#1\endcsname\relax
\typeout{** WARNING: IEEEtran.bst: No hyphenation pattern has been}%
\typeout{** loaded for the language `#1'. Using the pattern for}%
\typeout{** the default language instead.}%
\else
\language=\csname l@#1\endcsname
\fi
#2}}
\providecommand{\BIBdecl}{\relax}
\BIBdecl

\bibitem{Chen2021}
X.~Chen, D.~W.~K. Ng, W.~Yu, E.~G. Larsson, N.~Al-Dhahir, and R.~Schober,
  ``Massive access for {5G} and beyond,'' \emph{{IEEE} J. Select. Areas
  Commun.}, vol.~39, no.~3, pp. 615--637, 2021.

\bibitem{Callebaut2021}
G.~Callebaut, S.~Gunnarsson, A.~P. Guevara, A.~J. Johansson, L.~Van Der~Perre,
  and F.~Tufvesson, ``Experimental exploration of unlicensed sub-{GHz} massive
  {MIMO} for massive internet-of-things,'' \emph{{IEEE} Open Jour. Commun.
  Soc.}, vol.~2, pp. 2195--2204, 2021.

\bibitem{Yan2020}
H.~Yan and H.~Yang, ``Pilot length and channel estimation for massive {MIMO}
  {IoT} systems,'' \emph{{IEEE} Trans. Veh. Technol.}, vol.~69, no.~12, pp.
  15\,532--15\,544, 2020.

\bibitem{Lee2020}
B.~M. Lee and H.~Yang, ``Massive {MIMO} with massive connectivity for
  industrial internet of things,'' \emph{{IEEE} Trans. Ind. Electron.},
  vol.~67, no.~6, pp. 5187--5196, 2020.

\bibitem{Senel2018}
K.~{Senel} and E.~G. {Larsson}, ``Grant-free massive {MTC}-enabled massive
  {MIMO}: A compressive sensing approach,'' \emph{{IEEE} Trans. Commun.},
  vol.~66, no.~12, pp. 6164--6175, 2018.

\bibitem{Marzetta2010}
T.~L. Marzetta, ``Noncooperative cellular wireless with unlimited numbers of
  base station antennas,'' \emph{{IEEE} Trans. Wireless Commun.}, vol.~9,
  no.~11, pp. 3590--3600, 2010.

\bibitem{Hoydis2013}
J.~Hoydis, S.~ten Brink, and M.~Debbah, ``Massive {MIMO} in the {UL/DL} of
  cellular networks: How many antennas do we need?'' \emph{{IEEE} J. Select.
  Areas Commun.}, vol.~31, no.~2, pp. 160--171, 2013.

\bibitem{Ngo2013}
H.~Q. Ngo, E.~G. Larsson, and T.~L. Marzetta, ``Energy and spectral efficiency
  of very large multiuser {MIMO} systems,'' \emph{{IEEE} Trans. Commun.},
  vol.~61, no.~4, pp. 1436--1449, 2013.

\bibitem{Bockelmann2016}
C.~{Bockelmann}, N.~{Pratas}, H.~{Nikopour}, K.~{Au}, T.~{Svensson},
  C.~{Stefanovic}, P.~{Popovski}, and A.~{Dekorsy}, ``Massive machine-type
  communications in {5G}: physical and {MAC}-layer solutions,'' \emph{{IEEE}
  Commun. Mag.}, vol.~54, no.~9, pp. 59--65, 2016.

\bibitem{Cheng2021}
Y.~Cheng, L.~Liu, and L.~Ping, ``Orthogonal {AMP} for massive access in
  channels with spatial and temporal correlations,'' \emph{{IEEE} J. Select.
  Areas Commun.}, vol.~39, no.~3, pp. 726--740, 2021.

\bibitem{Liu2018}
L.~Liu and W.~Yu, ``Massive connectivity with massive {MIMO}—part {I}: Device
  activity detection and channel estimation,'' \emph{{IEEE} Trans. Signal
  Processing}, vol.~66, no.~11, pp. 2933--2946, 2018.

\bibitem{Le2021}
M.~T.~P. Le, L.~Sanguinetti, E.~Björnson, and M.-G.~D. Benedetto,
  ``Code-domain {NOMA} in massive {MIMO}: When is it needed?'' \emph{{IEEE}
  Trans. Veh. Technol.}, vol.~70, no.~5, pp. 4709--4723, 2021.

\bibitem{Larsson2014}
E.~G. {Larsson}, O.~{Edfors}, F.~{Tufvesson}, and T.~L. {Marzetta}, ``Massive
  {MIMO} for next generation wireless systems,'' \emph{{IEEE} Commun. Mag.},
  vol.~52, no.~2, pp. 186--195, 2014.

\bibitem{Bjornson2016}
E.~{Björnson}, E.~G. {Larsson}, and M.~{Debbah}, ``Massive {MIMO} for maximal
  spectral efficiency: How many users and pilots should be allocated?''
  \emph{{IEEE} Trans. Wireless Commun.}, vol.~15, no.~2, pp. 1293--1308, 2016.

\bibitem{Carvalho2017}
E.~D. {Carvalho}, E.~{Bjornson}, J.~H. {Sorensen}, P.~{Popovski}, and E.~G.
  {Larsson}, ``Random access protocols for massive {MIMO},'' \emph{{IEEE}
  Commun. Mag.}, vol.~55, no.~5, pp. 216--222, 2017.

\bibitem{Jing2018}
X.~Jing, M.~Li, H.~Liu, S.~Li, and G.~Pan, ``Superimposed pilot optimization
  design and channel estimation for multiuser massive {MIMO} systems,''
  \emph{{IEEE} Trans. Veh. Technol.}, vol.~67, no.~12, pp. 11\,818--11\,832,
  2018.

\bibitem{Lago2020}
L.~A. Lago, Y.~Zhang, N.~Akbar, Z.~Fei, N.~Yang, and Z.~He, ``Pilot
  decontamination based on superimposed pilots assisted by time-multiplexed
  pilots in massive {MIMO} networks,'' \emph{{IEEE} Trans. Veh. Technol.},
  vol.~69, no.~1, pp. 405--417, 2020.

\bibitem{Van_Chien2018}
T.~Van~Chien, E.~Björnson, and E.~G. Larsson, ``Joint pilot design and uplink
  power allocation in multi-cell massive {MIMO} systems,'' \emph{{IEEE} Trans.
  Wireless Commun.}, vol.~17, no.~3, pp. 2000--2015, 2018.

\bibitem{Xu2019}
J.~Xu, P.~Zhu, J.~Li, and X.~You, ``Deep learning-based pilot design for
  multi-user distributed massive {MIMO} systems,'' \emph{{IEEE} Wireless
  Commun. Lett.}, vol.~8, no.~4, pp. 1016--1019, 2019.

\bibitem{Sanguinetti2020}
L.~{Sanguinetti}, E.~{Björnson}, and J.~{Hoydis}, ``Toward massive {MIMO} 2.0:
  Understanding spatial correlation, interference suppression, and pilot
  contamination,'' \emph{{IEEE} Trans. Commun.}, vol.~68, no.~1, pp. 232--257,
  2020.

\bibitem{Muppirisetty2015}
L.~S. {Muppirisetty}, H.~{Wymeersch}, J.~{Karout}, and G.~{Fodor},
  ``Location-aided pilot contamination elimination for massive {MIMO}
  systems,'' in \emph{Proc. IEEE Global Telecommun. Conf.}, 2015, pp. 1--5.

\bibitem{Li2018}
P.~{Li}, Y.~{Gao}, Z.~{Li}, and D.~{Yang}, ``User grouping and pilot allocation
  for spatially correlated massive {MIMO} systems,'' \emph{{IEEE} Acc.},
  vol.~6, pp. 47\,959--47\,968, 2018.

\bibitem{Zhu2015}
X.~Zhu, L.~Dai, and Z.~Wang, ``Graph coloring based pilot allocation to
  mitigate pilot contamination for multi-cell massive {MIMO} systems,''
  \emph{{IEEE} Commun. Lett.}, vol.~19, no.~10, pp. 1842--1845, 2015.

\bibitem{Dao2018}
H.~T. {Dao} and S.~{Kim}, ``Vertex graph-coloring-based pilot assignment with
  location-based channel estimation for massive {MIMO} systems,'' \emph{{IEEE}
  Acc.}, vol.~6, pp. 4599--4607, 2018.

\bibitem{Ahmed2019}
\BIBentryALTinterwordspacing
A.~S. Al-hubaishi, N.~K. Noordin, A.~Sali, S.~Subramaniam, and
  A.~Mohammed~Mansoor, ``An efficient pilot assignment scheme for addressing
  pilot contamination in multicell massive {MIMO} systems,''
  \emph{Electronics}, vol.~8, no.~4, 2019. [Online]. Available:
  \url{https://www.mdpi.com/2079-9292/8/4/372}
\BIBentrySTDinterwordspacing

\bibitem{You2015}
L.~{You}, X.~{Gao}, X.~{Xia}, N.~{Ma}, and Y.~{Peng}, ``Pilot reuse for massive
  {MIMO} transmission over spatially correlated {R}ayleigh fading channels,''
  \emph{{IEEE} Trans. Wireless Commun.}, vol.~14, no.~6, pp. 3352--3366, 2015.

\bibitem{Herdin2005}
M.~{Herdin}, N.~{Czink}, H.~{Ozcelik}, and E.~{Bonek}, ``Correlation matrix
  distance, a meaningful measure for evaluation of non-stationary {MIMO}
  channels,'' in \emph{Proc. IEEE Veh. Technol. Conf.}, vol.~1, 2005, pp.
  136--140 Vol. 1.

\bibitem{Nguyen2021}
T.~H. Nguyen, T.~V. Chien, H.~Q. Ngo, X.~N. Tran, and E.~Björnson, ``Pilot
  assignment for joint uplink-downlink spectral efficiency enhancement in
  massive {MIMO} systems with spatial correlation,'' \emph{{IEEE} Trans. Veh.
  Technol.}, vol.~70, no.~8, pp. 8292--8297, 2021.

\bibitem{Ma2019}
J.~Ma, S.~Zhang, H.~Li, F.~Gao, and S.~Jin, ``Sparse bayesian learning for the
  time-varying massive {MIMO} channels: Acquisition and tracking,''
  \emph{{IEEE} Trans. Commun.}, vol.~67, no.~3, pp. 1925--1938, 2019.

\bibitem{Ribeiro2020}
L.~{Ribeiro}, M.~{Leinonen}, H.~{Djelouat}, and M.~{Juntti}, ``Channel charting
  for pilot reuse in {mMTC} with spatially correlated {MIMO} channels,'' in
  \emph{Proc. IEEE Global Telecommun. Conf. Worksh.}, 2020, pp. 1--6.

\bibitem{Ribeiro2021}
L.~{Ribeiro}, M.~{Leinonen}, H.~{Al-Tous}, O.~{Tirkkonen}, and M.~{Juntti},
  ``Exploiting spatial correlation for pilot reuse in single-cell {mMTC},'' in
  \emph{Proc. IEEE Int. Symp. Pers., Indoor, Mobile Radio Commun.}, 2021.

\bibitem{Studer2018}
C.~{Studer}, S.~{Medjkouh}, E.~{Gonultaş}, T.~{Goldstein}, and O.~{Tirkkonen},
  ``{Channel Charting}: Locating users within the radio environment using
  channel state information,'' \emph{{IEEE} Acc.}, vol.~6, pp.
  47\,682--47\,698, 2018.

\bibitem{Ribeiro2022}
L.~Ribeiro, M.~Leinonen, I.~Rathnayaka, H.~Al-Tous, and M.~Juntti, ``Channel
  charting aided pilot allocation in multi-cell massive {MIMO} {mMTC}
  networks,'' in \emph{Proc. IEEE Works. on Sign. Proc. Adv. in Wirel. Comms.},
  2022, pp. 1--5.

\bibitem{Bjornson2017}
\BIBentryALTinterwordspacing
E.~Björnson, J.~Hoydis, and L.~Sanguinetti, ``Massive {MIMO} networks:
  Spectral, energy, and hardware efficiency,'' \emph{Found. Trends Signal
  Proc.}, vol.~11, no. 3-4, pp. 154--655, 2017. [Online]. Available:
  \url{http://dx.doi.org/10.1561/2000000093}
\BIBentrySTDinterwordspacing

\bibitem{chen2018sparse}
Z.~Chen, F.~Sohrabi, and W.~Yu, ``Sparse activity detection for massive
  connectivity,'' \emph{{IEEE} Trans. Signal Processing}, vol.~66, no.~7, pp.
  1890--1904, 2018.

\bibitem{Djelouat2020Joint}
H.~{Djelouat}, M.~{Leinonen}, L.~{Ribeiro}, and M.~{Juntti}, ``Joint user
  identification and channel estimation via exploiting spatial channel
  covariance in m{MTC},'' \emph{{IEEE} Wireless Commun. Lett.}, pp. 1--1, 2021.

\bibitem{Morales2019}
J.~{García-Morales}, G.~{Femenias}, and F.~{Riera-Palou}, ``Higher order
  sectorization in {FFR}-aided {OFDMA} cellular networks: Spectral- and
  energy-efficiency,'' \emph{{IEEE} Acc.}, vol.~7, pp. 11\,127--11\,139, 2019.

\bibitem{Yu2021}
H.~Yu, L.~You, W.~Wang, and X.~Yi, ``Active channel sparsification for uplink
  massive {MIMO} with uniform planar array,'' \emph{{IEEE} Trans. Wireless
  Commun.}, vol.~20, no.~9, pp. 6018--6032, 2021.

\bibitem{TseBook}
D.~{Tse} and P.~{Viswanath}, \emph{Fundamentals of Wireless
  Communication}.\hskip 1em plus 0.5em minus 0.4em\relax Cambridge University
  Press, 2005.

\bibitem{Goldsmith2005}
A.~{Goldsmith}, \emph{Wireless Communications}.\hskip 1em plus 0.5em minus
  0.4em\relax Cambridge University Press, 2005.

\bibitem{TR36873}
3GPP, ``Technical specification group radio access network; study on {3D}
  channel model for {LTE},'' {V12.7.0}.

\bibitem{Kay1993}
S.~M. {Kay}, \emph{Fundamentals of Statistical Signal Processing: Estimation
  Theory}.\hskip 1em plus 0.5em minus 0.4em\relax Prentice Hall International,
  1993.

\bibitem{Yin2013}
H.~{Yin}, D.~{Gesbert}, M.~{Filippou}, and Y.~{Liu}, ``A coordinated approach
  to channel estimation in large-scale multiple-antenna systems,'' \emph{{IEEE}
  J. Select. Areas Commun.}, vol.~31, no.~2, pp. 264--273, 2013.

\bibitem{Kolen1996}
A.~W.~J. Kolen and J.~K. Lenstra, ``Combinatorics in operations research,'' in:
  {\it Handbook of Combinatorics}, R.L. Graham, M. Grötschel, L. Lovász
  (Eds.), Cambridge, MA, USA.: MIT Press, ch. 35, p. 1875–1910, 1996.

\bibitem{Vredeveld2003}
\BIBentryALTinterwordspacing
T.~Vredeveld and J.~K. Lenstra, ``On local search for the generalized graph
  coloring problem,'' \emph{Op. Res. Lett.}, vol.~31, no.~1, pp. 28--34, 2003.
  [Online]. Available:
  \url{https://www.sciencedirect.com/science/article/pii/S0167637702001657}
\BIBentrySTDinterwordspacing

\bibitem{Beyer1999}
K.~Beyer, J.~Goldstein, R.~Ramakrishnan, and U.~Shaft, ``When is ``nearest
  neighbor'' meaningful?'' in \emph{Database Theory --- ICDT'99}, C.~Beeri and
  P.~Buneman, Eds.\hskip 1em plus 0.5em minus 0.4em\relax Berlin, Heidelberg:
  Springer Berlin Heidelberg, 1999, pp. 217--235.

\bibitem{Zimek2012}
\BIBentryALTinterwordspacing
A.~Zimek, E.~Schubert, and H.-P. Kriegel, ``A survey on unsupervised outlier
  detection in high-dimensional numerical data,'' \emph{Statistical Analysis
  and Data Mining: The ASA Data Science Journal}, vol.~5, no.~5, pp. 363--387,
  2012. [Online]. Available:
  \url{https://onlinelibrary.wiley.com/doi/abs/10.1002/sam.11161}
\BIBentrySTDinterwordspacing

\bibitem{Lei2019}
E.~{Lei}, O.~{Castañeda}, O.~{Tirkkonen}, T.~{Goldstein}, and C.~{Studer},
  ``Siamese neural networks for wireless positioning and channel charting,'' in
  \emph{Proc. Annual Allerton Conf. Commun., Contr., Computing}, 2019, pp.
  200--207.

\bibitem{Huang2019}
P.~{Huang}, O.~{Castañeda}, E.~{Gönültaş}, S.~{Medjkouh}, O.~{Tirkkonen},
  T.~{Goldstein}, and C.~{Studer}, ``Improving channel charting with
  representation-constrained autoencoders,'' in \emph{Proc. IEEE Works. on
  Sign. Proc. Adv. in Wirel. Comms.}, 2019, pp. 1--5.

\bibitem{Agostini2020}
P.~{Agostini}, Z.~{Utkovski}, and S.~{Stańczak}, ``Channel charting: an
  {E}uclidean distance matrix completion perspective,'' in \emph{Proc. IEEE
  Int. Conf. Acoust., Speech, Signal Processing}, 2020, pp. 5010--5014.

\bibitem{Bishop2006}
C.~Bishop, \emph{Pattern Recognition and Machine Learning}.\hskip 1em plus
  0.5em minus 0.4em\relax Springer-Verlag New York, 2006.

\bibitem{Williams2002}
C.~K.~I. Williams, ``On a connection between kernel {PCA} and metric
  multidimensional scaling,'' \emph{Machine Learning}, vol.~46, pp. 11--19,
  2002.

\bibitem{Tenenbaum2000}
\BIBentryALTinterwordspacing
J.~B. Tenenbaum, V.~d. Silva, and J.~C. Langford, ``A global geometric
  framework for nonlinear dimensionality reduction,'' \emph{Science}, vol. 290,
  no. 5500, pp. 2319--2323, 2000. [Online]. Available:
  \url{https://science.sciencemag.org/content/290/5500/2319}
\BIBentrySTDinterwordspacing

\bibitem{Maaten2009}
L.~Van Der~Maaten, E.~Postma, J.~Van~den Herik \emph{et~al.}, ``Dimensionality
  reduction: a comparative,'' \emph{J Mach Learn Res}, vol.~10, no. 66-71,
  p.~13, 2009.

\bibitem{Dijkstra}
E.~W. Dijkstra, ``A note on two problems in connexion with graphs,''
  \emph{Numer. Math.}, vol.~1, no.~1, p. 269–271, dec 1959.

\bibitem{Floyd}
R.~W. Floyd, ``Algorithm 97: Shortest path,'' \emph{Commun. ACM}, vol.~5,
  no.~6, p. 345, jun 1962.

\bibitem{Milton2003}
J.~Milton and J.~Arnold, \emph{Introduction to Probability and Statistics:
  Principles and Applications for Engineering and the Computing
  Sciences}.\hskip 1em plus 0.5em minus 0.4em\relax McGraw-Hill, 2003.

\bibitem{Ponnada2019}
T.~Ponnada, H.~Al-Tous, O.~Tirkkonen, and C.~Studer, ``An out-of-sample
  extension for wireless multipoint channel charting,'' in \emph{Cognitive
  Radio-Oriented Wireless Networks}, 2019, pp. 208--217.

\bibitem{Silva2003}
\BIBentryALTinterwordspacing
V.~Silva and J.~Tenenbaum, ``Global versus local methods in nonlinear
  dimensionality reduction,'' in \emph{Advances in Neural Information
  Processing Systems}, S.~Becker, S.~Thrun, and K.~Obermayer, Eds.,
  vol.~15.\hskip 1em plus 0.5em minus 0.4em\relax MIT Press, 2003. [Online].
  Available:
  \url{https://papers.nips.cc/paper/2002/file/5d6646aad9bcc0be55b2c82f69750387-Paper.pdf}
\BIBentrySTDinterwordspacing

\bibitem{Tzeng2008}
J.~Tzeng, H.~H.-S. Lu, and W.-H. Li, ``Multidimensional scaling for large
  genomic data sets,'' \emph{BMC bioinformatics}, vol.~9, no.~1, pp. 1--17,
  2008.

\bibitem{Pedersen2000}
K.~Pedersen, P.~Mogensen, and B.~Fleury, ``A stochastic model of the temporal
  and azimuthal dispersion seen at the base station in outdoor propagation
  environments,'' \emph{{IEEE} Trans. Veh. Technol.}, vol.~49, no.~2, pp.
  437--447, 2000.

\bibitem{Ledoit2004}
O.~Ledoit and M.~Wolf, ``A well-conditioned estimator for large-dimensional
  covariance matrices,'' \emph{Journal of Multivariate Analysis}, vol.~88,
  no.~2, pp. 365--411, 2004.

\end{thebibliography}
\mybibliography

\end{document}